*nanomaterials*

MDPI

# Exchange Bias Effects in Iron Oxide-Based Nanoparticle Systems

**Manh-Huong Phan [1,\*], Javier Alonso [1,2], Hafsa Khurshid [1], Paula Lampen-Kelley [1], Sayan Chandra [1], Kristen Stojak Repa [1], Zohreh Nemati [1], Raja Das [1], Òscar Iglesias [3] and Hariharan Srikanth [1,\*]**

[1] Department of Physics, University of South Florida, Tampa, FL 33620, USA; jalonsomasa@gmail.com (J.A.); hafsamer@gmail.com (H.K.); lampenpa@gmail.com (P.L.-K.); sayan.chandra@gmail.com (S.C.); kstojak@mail.usf.edu (K.S.R.); zohrehnemati@mail.usf.edu (Z.N.); rajadas@mail.usf.edu (R.D.)

[2] BCMaterials Edificio No. 500, Parque Tecnológico de Vizcaya, 48160 Derio, Spain

[3] Department of Condensed Matter Physics and Institute of Nanoscience and Nanotechnology (IN2UB), University of Barcelona, Av. Diagonal 647, 08028 Barcelona, Spain; oscar@ffn.ub.es

\* Correspondence: phanm@usf.edu (M.-H.P.); sharihar@usf.edu (H.S.);
Tel.: +1-813-974-4322 (M.-H.P.); +1-813-974-2467 (H.S.); Fax:

**Abstract:** The exploration of exchange bias (EB) on the nanoscale provides a novel approach to improving the anisotropic properties of magnetic nanoparticles for prospective applications in nanospintronics and nanomedicine. However, the physical origin of EB is not fully understood. Recent advances in chemical synthesis provide a unique opportunity to explore EB in a variety of iron oxide-based nanostructures ranging from core/shell to hollow and hybrid composite nanoparticles. Experimental and atomistic Monte Carlo studies have shed light on the roles of interface and surface spins in these nanosystems. This review paper aims to provide a thorough understanding of the EB and related phenomena in iron oxide-based nanoparticle systems, knowledge of which is essential to tune the anisotropic magnetic properties of exchange-coupled nanoparticle systems for potential applications.

**Keywords:** iron oxide; nanostructure; exchange bias; spintronics; biomedicine

## 1. Introduction

Since the exchange bias (EB) phenomenon was first reported in ferromagnetic/antiferromagnetic (FM/AFM) Co/CoO core/shell nanoparticles in 1956 by Meiklejohn and Bean [1], it has generated a growing interest in the scientific community due to its current and potential technological value in spin valves, MRAM circuits, magnetic tunnel junctions, and spintronic devices [2–8]. EB is currently exploited to pin the magnetically hard reference layer in spin-valve read-back heads and MRAM memory circuits, as well as to increase the thermal stability of fine magnetic particles in advanced disk media. According to the Web of Science, the number of articles highlighting EB in the last 25 years has increased from fewer than 20 in the 1990s to an average of 300 articles per year in the last decade (see Figure 1). The citation of such articles has undergone a similar exponential increase to over 8000 citations in the year of 2015. This trend is expected to continue as industrial demand fuels renewed interest in exchange-biased nanostructures [7,8].

Despite extensive investigations of EB in a wide range of magnetic systems for more than 50 years [1–8], the physical origin of the phenomena is still poorly understood [5–8]. EB is most typically manifested as a shift in the magnetic hysteresis loop along the magnetic field axis and as an enhancement of the coercive field ($Hc$) when a ferromagnetic (FM)/antiferromagnetic (AFM) binary system is cooled in a field from above the AFM Néel temperature ($T_N$) [3,9,10]. In FM/AFM thin film systems and core/shell nanoparticles, EB was initially ascribed to exchange coupling between AFM and FM moments at the interface, the strength of which depends strongly on the relative anisotropy energies of the layers [2,3]. A recent comprehensive study of exchange biased polycrystalline FM/AFM films by O'Grady et al. points instead to the existence of disordered interfacial spins or "spin clusters", analogous to a spin glass at the interface between the FM and AFM layers [6]. These clusters are thought to transmit the anisotropy from the AFM layer to the FM layer, allowing for interactions via the exchange coupling to the AFM and FM layers that gives rise to the coercivity of the FM layer. In FeNi/CoO FM/AFM films, Berkowitz et al. estimate that ~75% of interfacial spins belong to magnetically hard particles (e.g., $CoFe_2O_4$) which are exchange-coupled to CoO and are responsible for the EB and coercivity enhancement by virtue of their exchange coupling to the FeNi layer [11]. The interfacial spins within the magnetically soft FeNi phase do not contribute to the EB. Ali et al. have made an elegant demonstration of how EB arises in Co/CuMn thin film interfaces formed between conventional ferromagnets and spin glasses (SG) [12]. These studies highlight the key observation, particularly relevant to many nanoparticle systems, that the presence of a conventional FM/AFM interface is not necessary to induce EB in all cases (see Figure 2 and Table 1). Experimental observations of EB in various types of interfaces, such as ferromagnet/ferrimagnet (FM/FI) [13], soft ferromagnet/hard ferromagnet (soft FM/hard FM) [14], ferrimagnet/antiferromagnet (FI/AFM) [15], ferrimagnet/ferrimagnet (FI/FI) [16], antiferromagnet/diluted ferromagnetic semiconductor (AFM/DMS) [17], have reflected its diverse origin.

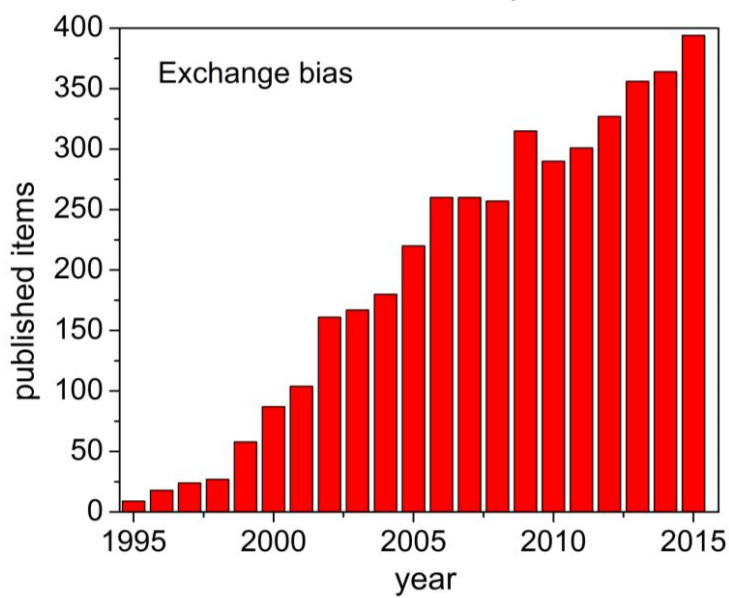

**Figure 1.** Number of published articles per year in the exchange bias research field. The data has been collected from Web of Science with "exchange bias" as keyword.

EB has also been reported in small ferrite particles with disordered surface spins, such as $NiFe_2O_4$ [18], $\gamma$-$Fe_2O_3$ [19], and $CoFe_2O_4$ [20]. The origin of this phenomenon is the fraction of surface spins with decreased co-ordination (and thus weaker bonding) increasing with a decrease in particle size. These disordered spins can take on a number of configurations, one of which can be chosen by field-cooling the particle to induce an EB [18–20]. The degree of disorder of the surface spins of the shell to which the ferrimagnetically ordered spins of the core couple is thought to be crucial for achieving EB in these nanoparticles. However, the underlying nature of spin ordering and the EB-related phenomena in ferrite nanoparticles has remained elusive, primarily due to the complex interplay between particle-size effects, inter-particle interactions and the random distribution of anisotropy axes throughout a given system [5].

Iron oxide (magnetite or maghemite) nanoparticles with desirable magnetic properties and biocompatibility have been extensively studied over the past several years for potential applications in nanomedicine [21–23]. Recently, iron/iron oxide core/shell nanoparticle systems have been exploited for such applications as the combination of the high magnetization of the core (Fe) and the chemical stability and biocompatibility of the shell ($Fe_3O_4$ or $\gamma$-$Fe_2O_3$) leads to more suitable overall properties than either material alone [24–26]. These systems also provide excellent models for probing the roles played by interface and surface spins and their impacts on EB in exchange-coupled nanostructures, inspiring a large body of work on core/shell Fe/$\gamma$-$Fe_2O_3$ [27–29], Fe/$Fe_3O_4$ [30,31], FeO/$Fe_3O_4$/$\gamma$-$Fe_2O_3$ [32–34], $Fe_3O_4$/$\gamma$-$Fe_2O_3$ [35], $\gamma$-$Fe_2O_3$/CoO [36], and Au/$Fe_3O_4$ particles [37–39]. Oxidization-driven migration of metal atoms from the core to the shell of iron/iron oxide nanoparticle systems has been shown to occur via the Kirkendall effect, leading to a morphological transformation into hollow nanoparticles, where additional (inner) surface area strongly affects magnetic properties including EB [40–44].

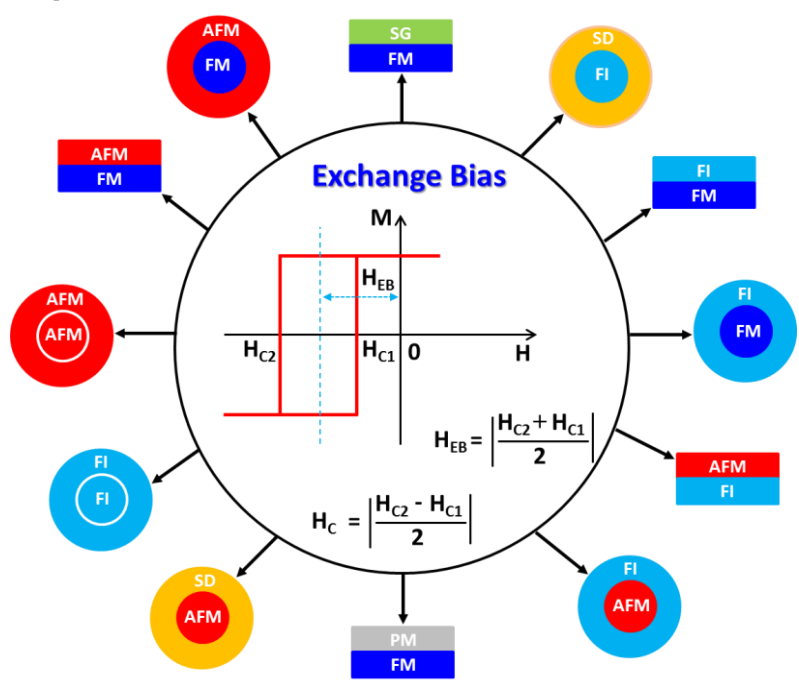

**Figure 2.** Schematic of the exchange bias (EB) phenomenon as a shift in the magnetic hysteresis loop at a low temperature when the sample is cooled in the presence of a magnetic field from a high temperature, well above the Néel or blocking temperature. The EB effects have been reported in a wide range of magnetic systems with different types of interfaces. FM, FI, AFM, SG, SD, and PM stand for ferromagnet, ferrimagnet, antiferromagnet, spin glass, spin disorder, and paramagnet, respectively.

In this review paper, we attempt to elucidate the roles of interface and surface spins in exchange biased iron oxide-based nanosystems through experimental and atomistic Monte Carlo studies that we have performed over the last several years. We first present a brief overview of crystal structures, electronic configurations, and size-dependent magnetic properties of magnetite ($Fe_3O_4$) and maghemite ($\gamma$-$Fe_2O_3$), which form the basis for a number of novel nanostructures. Subtle differences in the magnetic structure between the two materials lead to significant phenomenological variation in the EB effect of magnetite-based vs. maghemite-based nanoparticle systems. The physics of interface and surface spins and their impacts on EB in iron oxide-based nanostructures in various forms—solid single-component, core/shell, hollow, and hybrid composite nanoparticles—are discussed in detail. Finally, we present concluding remarks and an outlook for future research in this exciting field.

## 2. Fundamental Aspects of Magnetite ($Fe_3O_4$) and Maghemite ($\gamma$-$Fe_2O_3$)

## 2.1. Crystal Structure and Electronic Configuration

Magnetite and maghemite are ferrimagnetic iron oxides with a similar cubic structure [45] not distinguishable by standard resolution X-ray diffraction (XRD) [46,47]. Magnetite ($Fe_3O_4$) exhibits a spinel structure with a unit cell comprised of eight cubic units with a lattice spacing of 8.39 Å. The unit cell contains 56 atoms, including 32 oxygen atoms, 16 $Fe^{3+}$ ions and 8 $Fe^{2+}$ ions. As shown in Figure 3a, there are 32 octahedral (B) and 64 tetrahedral (A) sites in the unit cell [48]. The $Fe^{2+}$ cations occupy 1/4 of the octahedral interstitial sties and $Fe^{3+}$ cations are split, with 1/4 on the octahedral sites and 1/8 on the tetrahedral sites. For this reason, magnetite's unit cell can be represented as $(Fe^{3+})_8[Fe^{2.5+}]_{16}O_{32}$, where the parentheses () designate tetrahedral sites and the brackets [] designate octahedral sites. This crystallographic configuration is denoted as "inverse spinel".

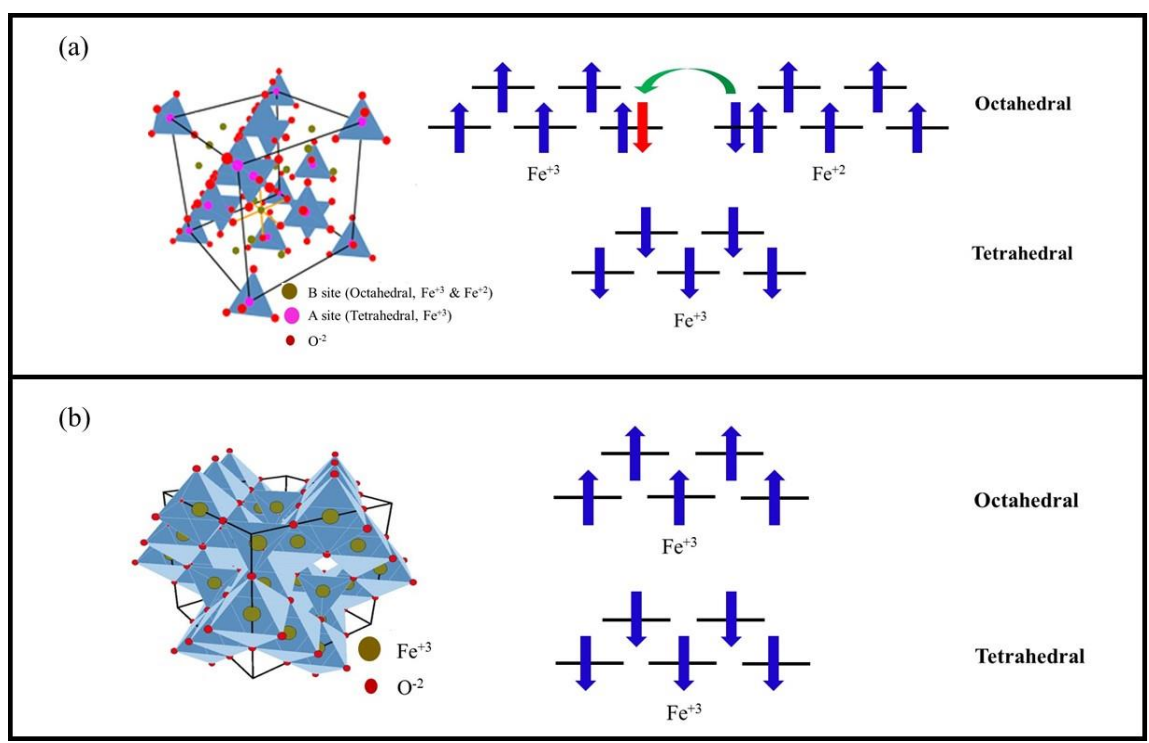

**Figure 3.** Crystal structures and spin configurations of (**a**) magnetite $Fe_3O_4$ and (**b**) maghemite $\gamma$-$Fe_2O_3$. (Courtesy of Aidin Lak, Reprinted with permission from [48]).

Maghemite [$\gamma$-$Fe_2O_3$] crystallizes in a similar cubic spinel structure with a tetragonal supercell and a lattice constant of 8.33 Å (Figure 3b) [48]. In contrast to magnetite, Fe occurs in a single oxidation state ($Fe^{3+}$) in maghemite, with $Fe^{3+}$ cations arbitrarily distributed in 16 octahedral and 8 tetrahedral interstitial sites [48]. The maghemite structure can be obtained by creating 8/3 vacancies out of the 24 Fe sites in the cubic unit cell of magnetite. There is experimental [49] and theoretical [50] evidence that $Fe^{3+}$ cations and vacancies tend to order in the octahedral sites, maximizing the homogeneity of the distribution and minimizing the electrostatic energy of the crystal. Therefore, the structure of maghemite can be approximated as a cubic unit cell with the composition $(Fe^{3+})_8[Fe^{3+}_{5/6} \square_{1/6}]_{16}O_{32}$.

Concerning the magnetic configuration in the case of magnetite, octahedrally coordinated $Fe^{3+}$ and $Fe^{2+}$ ions are coupled ferromagnetically through a double exchange mechanism. The electron whose spin is directed in the opposite direction of the others and colored red, can be exchanged between two octahedral coordination sites (Figure 3a). On the other hand, $Fe^{3+}$ ions in tetrahedral and octahedral sites are coupled antiferromagnetically via the oxygen atom, yielding a net zero magnetization in the $Fe^{3+}$ sublattice. The ferrimagnetic moment in magnetite thus arises from the unpaired spins of $Fe^{2+}$ in octahedral coordination. In maghemite, $Fe^{3+}$ ions on the tetrahedral and octahedral sites are coupled antiferromagnetically via the oxygen atom; this leaves unpaired octahedral $Fe^{3+}$ spins to contribute to the magnetization.

The saturation magnetization ($M_S$) values are quite similar for both compounds ($M_{SFe3O4} \cong 90$ emu/g and $M_{S\gamma\text{-}Fe2O3} \cong 84$ emu/g) [51]. A primary difference in the magnetic properties is the Curie temperature (893 K for maghemite and 858 K for magnetite). In addition, a first order magnetic/structural transition (the well-known Verwey transition) occurs at $T_v \sim 115$ K in bulk magnetite, but is absent in the case of maghemite [52,53]. $T_v$ has been reported to shift to lower temperatures in magnetite nanoparticle systems as particle size is decreased [54].

## 2.2. Size-Dependent Magnetic Properties

Reducing the size of a ferromagnetic system to the nanometer scale (below 100 nm) has been shown to strongly alter its magnetic properties [55,56]. As a ferromagnetic particle reaches a material-dependent critical domain size ($D_C$, the critical diameter), it contains a single magnetic domain [56,57]. As particle size becomes smaller than $D_C$, the formation of domain walls becomes energetically unfavorable and the single domain (SD) particles are obtained. Below $D_C$, the coercivity varies with the particle size as $H_C \sim D^6$. As particle size is decreased further below $D_C$, a second threshold $D_{sp}$ appears in which thermal fluctuations easily destabilize the ordered magnetic state and the system exhibits superparamagnetic behavior. For spherical $Fe_3O_4$ and $\gamma\text{-}Fe_2O_3$ particles, $D_C \cong 82$ nm and 91 nm, respectively [57]. It is worth noting that particles with significant shape anisotropy can remain single domain at much larger sizes than their spherical counterparts [56,57].

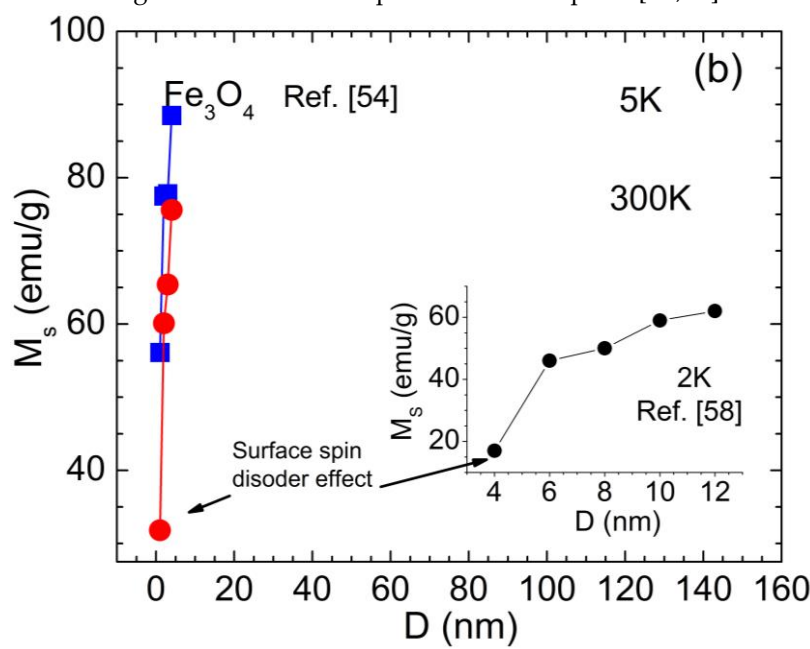

**Figure 4.** Particle diameter dependences of (**a**) coercive field ($H_C$) and (**b**) saturation magnetization ($M_S$) for spherical $Fe_3O_4$ nanoparticles and for (**c**) spherical $\gamma\text{-}Fe_2O_3$ nanoparticles [46,54,58].

The static magnetic properties of iron oxide MNPs depend strongly on particle size and shape [54–57]. Figure 4a illustrates the variation of $H_C$ with particle size for $Fe_3O_4$ MNP assemblies. Goya et al. [54] reported that with decreasing particle size from 150 nm to 4 nm, $H_C$ first decreased from 150 nm to 11.5 nm, but increased sharply for the smallest particles ($D = 4$ nm). Dutta et al. [58] observed a slight increase in $H_C$ as particle size was decreased from 12 nm to 6 nm, and a sudden increase for 4 nm $Fe_3O_4$ MNPs. In both cases, the enhancement of $H_C$ (Figure 4a) for 4 nm $Fe_3O_4$ MNPs is accompanied by a strong decrease of saturation magnetization $M_S$ (Figure 4b), a consequence of the strong surface spin disorder present in these MNPs. The fraction of spins on the surface of MNPs increases with decrease in particle size. It has been suggested that when the surface to volume ratio becomes sufficiently large, broken exchange bonds induce surface spin disorder, thus creating a core/shell structure comprised of a ferrite core with a shell of disordered spins [5]. By assuming a core/shell structure with a disordered shell of thickness $d$ (i.e., the magnetically dead layer) that does not contribute to $M_S$, the variation of $M_S$ with particle size $D$ can be expressed by [58]

$$M_S = M_0\left(1 - 2d/D\right)^3 \tag{1}$$

where $M_0$ is the saturation magnetization of the bulk material. Using this relationship, Dutta et al. [58] determined the thickness of the spin-disordered shell to be $d = 0.68$ nm for $Fe_3O_4$ MNPs with $D > 4$ nm and $d = 0.86$ nm for $Fe_3O_4$ MNPs with $D = 4$ nm. The remarkable increase in $d$ for $Fe_3O_4$ MNPs with $D = 4$ nm gives a natural explanation for the strong decrease in $M_S$ (Figure 4b) and increase in $H_C$ (Figure 4a).

A similar trend in the size-dependent magnetic properties has also been reported for $\gamma$-$Fe_2O_3$ MNPs [46]. The difference noted here is that the critical size below which $M_S$ decreases is drastically larger in spherical $\gamma$-$Fe_2O_3$ MNPs (Figure 4c) than spherical $Fe_3O_4$ MNPs (Figure 4b). This is in accordance with the experimental and theoretical studies that have shown that surface spins are more magnetically frustrated (more disordered) in $\gamma$-$Fe_2O_3$ MNPs than in $Fe_3O_4$ MNPs of the same size [5,47]. In $\gamma$-$Fe_2O_3$ MNPs, it has been reported that there exists a threshold size below and above which cation vacancies are respectively disordered and ordered [47]. As a result, both surface effects and bulk order-disorder effects are important in determining the low temperature magnetic behavior in small $\gamma$-$Fe_2O_3$ nanoparticles. Different degrees of vacancy ordering in $\gamma$-$Fe_2O_3$ nanoparticles, which are directly related to sample preparation methods, have been suggested to lead to some deviation in the magnetic parameters including EB fields, as reported in the literature [47,59].

In many cases, both magnetite and maghemite phases are present in ferrite MNPs. For the 4 nm particles, Dutta et al. [58] have revealed the presence of the $\gamma$-$Fe_2O_3$ phase rather than the $Fe_3O_4$ phase, suggesting a possible structural transformation from the magnetite to maghemite phase below ~5 nm. This hypothesis has been supported by a later study by Frison et al. [60] on the size-dependent magnetic properties of $Fe_3O_4$/$\gamma$-$Fe_2O_3$ MNPs in the 5–15 nm range. The authors showed that, upon size reduction from 15 to 5 nm, the volume fraction of the $Fe_3O_4$ phase decreased from ~60 to ~25 wt %, whereas the volume fraction of the $\gamma$-$Fe_2O_3$ phase increased respectively. Since the $M_S$ of bulk $\gamma$-$Fe_2O_3$ is smaller than that of bulk $Fe_3O_4$, a decrease in $M_S$ of $Fe_3O_4$/$\gamma$-$Fe_2O_3$ MNPs was observed as the $\gamma$-$Fe_2O_3$ to $Fe_3O_4$ ratio was increased [60], and the noted drop in $M_S$ for the $D = 4$ nm sample (Figure 4b) of Dutta et al. [58] may thus arise from a combination of surface spin disorder and the structural transformation into the $\gamma$-$Fe_2O_3$ phase.

The dynamic magnetic properties of iron oxide MNPs also depend sensitively on particle size and shape [54–58]. AC susceptibility measurements and analyses give important clues about the spin dynamics and the role of dipolar inter-particle interactions between MNPs forming clusters or arrays [55,56]. For a non-interacting MNP system, the frequency dependence of the blocking temperature ($T_B$) is predicted to follow a simple Arrhenius law, while for magnetically interacting MNP systems, the frequency dependence of $T_B$ is better represented by the Vogel-Fulcher law [56]. Goya et al. [54] fitted the AC susceptibility data of 5 nm $Fe_3O_4$ MNPs using the Arrhenius model and yielded the characteristic relaxation time, $\tau_0 = 0.9 \times 10^{-12}$ s, for this system. In the limit of superparamagnetic systems, $\tau_0$ decreases with decreasing particle size [54,56]. In the case of interacting MNP assemblies, both $\tau_0$ and $T_0$ (the characteristic temperature, which is considered as a measure of the strength of the dipolar interparticle interaction [56]) seem to increase with increasing particle size [56]. We note that both Vogel-Fulcher and Arrhenius models are useful for investigating the relaxation processes in non-interacting and weakly interacting MNP assemblies, but may not be appropriate for studying the spin dynamics of complex composite MNP systems, such as Au-$Fe_3O_4$ hybrid nanostructures [37], in which other types of interaction are dominant over dipolar inter-particle interactions. For the case of small ferrite MNPs with disordered surface spins, in addition to undergoing a superparamagnetic to blocked transition at $T_B$, the system enters a spin-glass-like state at a lower temperature (a so-called surface spin freezing temperature, $T_f$), below which a noticeable slowing down in the spin dynamics and a sizeable EB are observed [18–20].

### 3. Exchange Bias Effect in Single-Component Solid Nanoparticles

It has been suggested that in ferrite MNPs the disordered spins can take on a number of configurations, one of which can be chosen by field-cooling the particle to induce an EB effect [18–20]. The lowest energy configuration of surface spins in the zero-field cooled condition of a spherical particle is the one in which the spins point in the radial direction from the particle. The energy required to rotate these spins contributes to the enhanced coercivity below $T_F$ as well as to "open", irreversible hysteresis up to high fields [5,19].

In case of spherical $Fe_3O_4$ MNPs, a Monte Carlo simulation study found an EB effect for particle sizes less than 2.5 nm, where the surface anisotropy ($K_S$) resulting from disordered surface spins is assumed to be large compared to the core cubic magnetocrystalline anisotropy ($K_C$) [61]. More information on Monte Carlo Simulations can be found at the end of the article, in Appendix A. No EB is observed for spherical $Fe_3O_4$ MNPs with larger diameters. This prediction is in good agreement with the experimental study of Goya et al. [54] that reported the absence of EB in spherical $Fe_3O_4$ MNPs with particle sizes as small as 5 nm. Small-angle neutron scattering experiments with polarization analysis on spherical $Fe_3O_4$ MNPs of ~9 nm diameter have revealed a uniformly 90°-canted, magnetically active shell, rather than a shell of disordered spins [62]. This observation is key to explaining the absence of EB in ~9 nm $Fe_3O_4$ MNPs and other $Fe_3O_4$-based nanosystems [37,54]. It has been noted that while the magnetic properties of $Fe_3O_4$ MNPs are modified when their spherical form is transformed into other shapes like cubes and octopods, no EB has been reported for these morphologies [63]. This suggests that spin canting in the shell layer may not be strongly altered by varying particle shape. However, compacting 20 nm spherical $Fe_3O_4$ MNPs under high pressures ranging between 1 and 5 Gpa was found to introduce disorder in a layer of surface spins, giving rise to EB [63]. The formation of a core/shell magnetic structure during compaction lead to a field dependence of $T_F$ that follows the de Almeida–Thouless (AT) relationship. Interestingly, the authors have shown that there exists a critical cooling field ($H_{cri}$), above which both the surface spin-glass behavior and the EB effect abruptly disappear, and that both $H_E$ and $H_{cri}$ increase with the applied pressure (Figure 5a) [63].

In contrast to magnetite, the reported magnetic properties of $\gamma$-$Fe_2O_3$ MNPs, including EB, are sensitively dependent on particle size and shape [19,64–68]. As the particle size decreases from 21 to 7 nm, the thickness of the disordered shell of spins in maghemite increases from $d$ = 0.46 to 1.1 nm [64]. This increase in $d$ corresponds to a decrease in $M_S$ from 70 to 24 emu/g. We note that a sharp increase and decrease in $d$ and $M_S$ were observed for spherical $\gamma$-$Fe_2O_3$ MNPs with an average diameter just less than 10 nm. An EB effect was observed by Martinez et al. [64] at temperatures below a low temperature surface spin-glass-like transition ($T_F$ ~ 42 K) for ~10 nm $\gamma$-$Fe_2O_3$ MNPs with platelet-like shapes. The magnetic field dependence of $T_F$ followed the AT line, characteristic of many magnetic glassy systems. In addition to surface spin disorder and particle size effects, inter-particle interactions, often present in nanoparticle assemblies, could be a supplementary source of the magnetic frustration that produces a frozen collective state of the particle spins at low temperatures. To address this, Shendruk et al. [65] performed a systematic study of the static and dynamic magnetic properties, including EB, in ~7 nm spherical $\gamma$-$Fe_2O_3$ MNPs whose distances were well separated to an extent that the effect of inter-particle interactions was negligible. They have shown that the disordered shell spins are exchange-coupled to the spins of the ordered core, resulting in an EB loop shift, and that an unusual exponential-like decrease of the total $M_S$ of the $\gamma$-$Fe_2O_3$ MNPs with increasing temperature resulted from the surface spin disorder. A large EB effect has been reported for spherical $\gamma$-$Fe_2O_3$ MNPs of 4 nm diameter [58].

To theoretically probe the roles of surface anisotropy and disorder in these MNPs, Iglesias et al. [69–72] have performed Monte Carlo simulations of a single maghemite MNP with two different shapes (spherical vs. ellipsoidal). More information about these simulations can be found in Appendix A. They observed the formation of hedgehog-like spin structures due to increased surface anisotropy in case of the spherical $\gamma$-$Fe_2O_3$ MNPs [69]. The increase in surface spin disorder corresponds to the increased surface anisotropy and coercivity. As the nanoparticle with high surface anisotropy varies in shape from sphere to ellipse, the spins will point along the local radial direction and possess antiparallel orientations with the nearest neighbors on the other sublattice [71]. To illustrate these features, in Figure 5b we present the simulated M-H loops for a spherical γ-$Fe_2O_3$ MNP with different values of surface anisotropy constants ($k_S$) and snapshots of the equilibrium spin configurations attained after cooling from a high temperature to $T = 0$ K for $k_S = 10$ and 50 K (Figure 5c,d) [70]. It can be seen that the hysteresis loops become elongated with high values of the differential susceptibility, resembling those from frustrated or disordered systems, which are in good agreement with the experimental observations [18–20]. For a given size of a nanoparticle, the increase in surface spin disorder causes $k_S$ to increase, thus strengthening the elongated shape (Figure 5b). This model can be applied to quantify the increases in $H_E$ and $H_{cri}$ for the 20 nm compacted $Fe_3O_4$ MNPs, when the applied pressure was increased [63]. However, the model does not predict the freezing of the layer of surface spins into a spin-glass-like state, [69,70] which has been proposed to explain experimental observations [18–20].

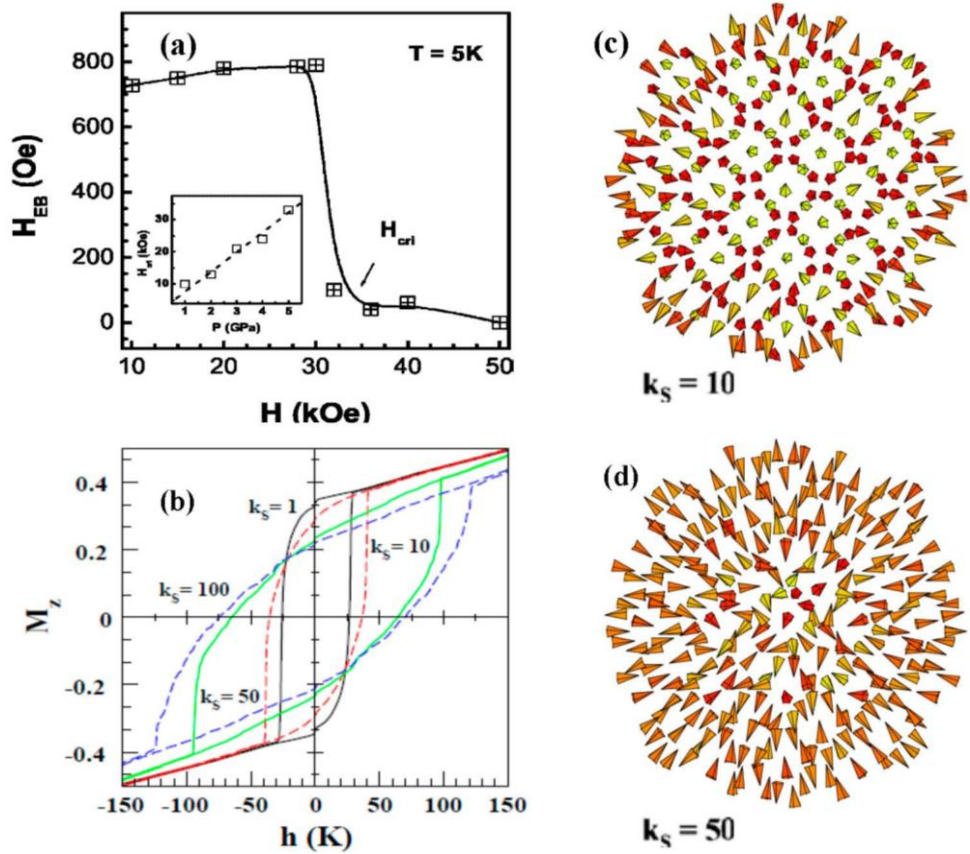

**Figure 5.** (**a**) Exchange bias field ($H_E$) for $Fe_3O_4$ 5-GPa-nanocompacts at 5 K as a function of the cooling field showing an abrupt decrease of $H_E$ when the cooling field exceeds a critical field, $H_{cri}$. Inset shows pressure dependence of $H_{cri}$. Reproduced with permission from [63]; (**b**) Hysteresis loops for a particle with $D = 5a$ ($D$: particle diameter; $a$: lattice constant) for different values of $k_S$ as indicated; (**c,d**) Schematic of the exchange bias snapshots of the corresponding $T = 0$ configurations of a spherical particle with $D = 5a$ obtained after cooling from a disordered state at high T for two different values of $k_S$ (small and large); Reproduced with permission from [70].

## 4. Exchange Bias Effect in Core/Shell Nanoparticles

Exchange bias has been observed in different types of iron oxide-based core/shell nanoparticles, such as Fe/γ-$Fe_2O_3$ [27–29], Fe/$Fe_3O_4$ [30,31], FeO/$Fe_3O_4$ [32–34], $Fe_3O_4$/γ-$Fe_2O_3$ [35], γ-$Fe_2O_3$/CoO [36], $MnFe_2O_4$/γ-$Fe_2O_3$, and $CoFe_2O_4$/γ-$Fe_2O_3$ [73]. The EB effect has been mainly associated with exchange anisotropy originating from exchange coupling between the shell and the core. The magnetic properties of the core/shell MNPs are dependent both on the core and the shell size, shape, composition, interface roughness, etc. [27–36,73–76]. These parameters can be easily tuned during the synthesis process and a rich variety of magnetic core/shell systems can be produced by a proper choice of the materials and dimensions of both the core and the shell. So far, various experimental methods have been employed to produce core/shell MNPs, such as chemical and thermal decomposition, ball milling, gas condensation, chemical vapor deposition, pulsed laser deposition, etc. [27,29,30,78,79]. In addition, different theoretical models and simulations have been developed in order to gain a better understanding of the role of different parameters that control the EB in these core/shell MNPs [5,40,80–83]. It has also been shown that both inter- and intra-particle interactions play important roles in the EB loop shift [73–75]. A training effect, namely, the decrease of EB field after consecutive field cooling hysteresis loop measurements, has also been observed in these MNPs, and is related to a decrease in the number of spins frozen along the magnetic field after each consecutive measurement [27].

It is generally accepted that core/shell nanoparticles are composed of different materials, and that the effective anisotropy, lattice strain, number of uncompensated spins etc. for the materials composing the core and shell are different [28]. This implies that the core and shell may have different responses to changes in temperature and magnetic field. If so, two important questions emerge: *Can the dynamic and static response of the core and shell be identified separately?* and *How does the EB depend on the magnetic states of the core and the shell?* To address these questions, we have performed a systematic study of the static and dynamic magnetic properties of Fe/$\gamma$-Fe$_2$O$_3$ core/shell nanoparticles (mean size, ~10 nm). These MNPs were prepared by high temperature reduction of iron pentacarbonyl in octadecene in the presence of olyelamine (OY) and trioctyl phosphine (TOP) [28,29], details of which can be found in Appendix B. As an example, in Figure 6 we present the structural and magnetic analysis of the Fe/$\gamma$-Fe$_2$O$_3$ nanoparticles. An HRTEM image (inset of Figure 6b) reveals the crystalline structure of both the core and shell with lattice spacing of the core and shell corresponding to (110) planes of bcc iron and (311) planes of fcc iron oxide, respectively. The Fe core is single crystalline, however, the iron oxide shell is composed of small crystallites that are oriented randomly [29]. An X-ray Absorption Near Edge Spectroscopy (XANES) study on these MNPs has recently confirmed that the shell is mostly formed by $\gamma$-Fe$_2$O$_3$ (rather than Fe$_3$O$_4$), while the core is composed of Fe [44]. Analysis of DC and AC susceptibility measurements has revealed that the particle dynamics critically slow down at $T_g$ ~ 68 K (Figure 6a), below which they exhibit memory effect in FC and ZFC protocols; this behavior is characteristic of a superspin glass (SSG) state. The field dependence of $T_B$ fits the AT line and shows two different linear responses in the low and high field regimes corresponding to the core and shell respectively (Figure 6b). The energy barrier distribution estimated from the temperature decay of isothermal remanent magnetization shows two maxima that mark the freezing temperatures of the core ($T_{f\text{-}cr}$ ~ 48 K) and shell ($T_{f\text{-}sh}$ ~ 21 K) [28]. The EB field has been observed to start developing from ~35 K, and as temperature decreases, $H_E$ increases slowly at first, followed by a rapid increase below ~21 K (Figure 6c). We find that at 35 K, the core is frozen with its spins aligned along the field and the shell begins to show a blocking behavior. Due to the slow dynamics of the blocked spins in the shell (<35 K) they behave as pinning centers, leading to the development of EB. This marks the onset of EB in the core/shell MNPs. Below 21 K ($\leq T_{f\text{-}sh}$), when the shell is completely frozen, the number of pinning centers increases due to enhanced exchange coupling between the core and the shell. Consistently, a rapid increase in $H_E$ is recorded just below $T_{f\text{-}sh}$. Accordingly, we suggest that in the case of core/shell nanoparticles, the onset of EB is associated with the blocking of spins in the shell while the core is in the frozen state. These findings are of practical importance in tailoring EB and its onset temperature by suitably choosing different materials for the core and shell that show blocking and freezing phenomena at a desired temperature range [28]. A schematic showing a temperature-dependent EB field trend and its association with the magnetic states of the core and shell for a core/shell nanoparticle system is proposed in Figure 6d. Apart from those, we note that while only a horizontal M-H shift is often observed in bilayered FM/AFM films, both horizontal and vertical M-H shifts have been reported for the core/shell MNPs (see inset of Figure 6c). This vertical M-H shift has been attributed to the presence of a number of frozen spins that cannot be reversed by the magnetic field [30]. The EB is directly related to the moment of irreversible spins, inversely related to the reversible spins, and depends linearly on the ratio of number of frozen spins to reversible spins [29–31]. Another interesting feature to note is the presence of a sharp change or a jump in the magnetization at low fields in the FC M-H loop (Figure 6c). This jump was initially attributed to the unidirectional alignment of frozen interfacial spins during the FC, which provided a maximum exchange coupling between the core and the shell for the case of Fe/Fe₃O₄ [31]. However, such a feature is still observed in hollow γ-Fe₂O₃ MNPs after the Fe core was removed from the Fe/γ-Fe₂O₃ core/shell structure [41,44].

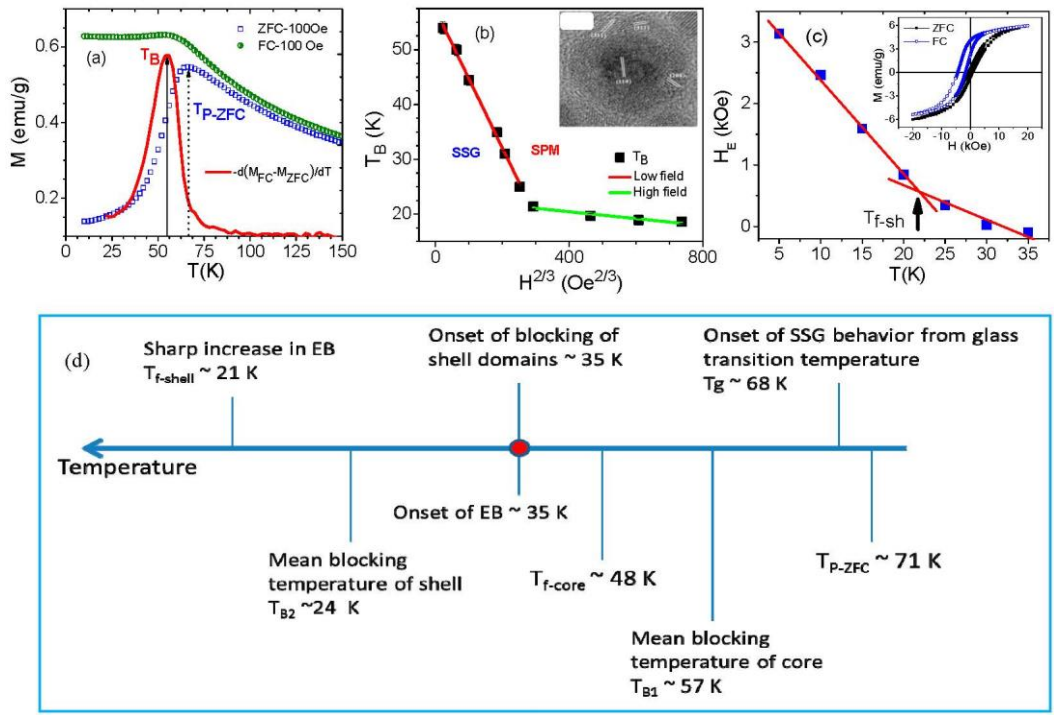

**Figure 6.** (**a**) ZFC/FC curve and the derivative $d[M_{FC}-M_{ZFC}]/dT$ indicating peak temperature of $M_{ZFC}$ ($T_{P\text{-}ZFC}$) and mean blocking temperature ($T_B$); (**b**) Two straight AT-line fits for low and high field regimes. Inset shows a TEM image of the Fe/γ-Fe₂O₃ core/shell nanoparticle; (**c**) Temperature dependence of exchange bias field ($H_E$). Inset shows the M-H loops at 5 K under FC of a 5 T field (open circles) and ZFC (filled circles) conditions [28]; (**d**) A schematic showing a temperature-dependent EB field trend and its association with the magnetic states of the core and shell for a core/shell nanoparticle system.

In a typical core/shell nanoparticle system, it is generally accepted that both interface and surface spins play important roles in triggering the EB effect. Indeed, studies have shown that both the interface and surface spins contribute to the EB effects in cases of Fe/γ-Fe₂O₃ [27–29], Fe₃O₄/γ-Fe₂O₃ [35], CoO/γ-Fe₂O₃ [36], MnFe₂O₄/γ-Fe₂O₃, and CoFe₂O₄/γ-Fe₂O₃ [73] core/shell MNPs. However, Ong et al. have revealed that the frozen spins at the interface between the core and the shell dominate the EB effect in the case of ~14 nm Fe/Fe₃O₄ core/shell MNPs [30,31]. This hypothesis has been supported by the observation of a largely reduced EB field in hollow Fe₃O₄ MNPs of ~4.5 nm thickness when the Fe core was removed from the core/shell structure [31]. The crucial role of the frozen interfacial spins has also been proposed for obtaining the large EB effects in FeO/Fe₃O₄ core/shell MNPs [32–34]. These varied observations raise a fundamental question of *how to decouple collective contributions to the EB from the interface and surface spins in such a core/shell nanoparticle system*? We have addressed this through a comparative study of the EB effect in Fe/γ-Fe₂O₃ core/shell MNPs with the same thickness of the γ-Fe₂O₃ shell (~2 nm) and the diameter of the Fe core varying from 4 nm to 11 nm (see TEM images of these MNPs in Figure 7a–c). Magnetic measurements have revealed that decreasing particle size significantly decreases the magnetization and increases the magnetic anisotropy (Figure 7d), suggesting an enhanced disordering effect of surface spins in the samples with reduced particle size [5]. To probe EB in these MNPs, the samples were cooled to 5 K from room temperature in the presence of a 5 T magnetic field. For all of the samples, $H_E$ decreases exponentially with temperature, becoming nearly null around 30 K, and reaching a maximum value of 3.5 kOe at 5 K for the smallest nanoparticles. In order to quantify the contribution to the EB from the interface spins for these MNPs, we have used the modified Meiklejohn and Bean (MB) model, which was initially developed for EB in AFM/FM coupled thin films [1]. In case of a core/shell FM/FI nanoparticle system, $H_E$ can be expressed as

$$H_{EB} = 2 \frac{nJ_{ex}S_{FM}S_{FIM}}{a^2 M_{FM} t_{FM}} = \frac{\Delta E}{2 M_{FM} t_{FM}} \qquad (2)$$

where $\Delta E$ is the interfacial exchange energy density needed to reverse the frozen spins, $J_{ex}$ is the interfacial exchange constant. $S_{FM}$ and $S_{FI}$ represent individual spin moments of the FM core and the FI shell, respectively. $M_{FM}$ and $t_{FM}$ are the saturation magnetization and effective thickness of the ferromagnetic layer, and $n/a^2$ is the number of exchange-coupled bonds across the interface per unit area. In the case of our Fe/$\gamma$-Fe$_2$O$_3$ nanoparticles, the decrease in particle size is expected to vary the interaction area between the Fe core and the $\gamma$-Fe$_2$O$_3$ shell within the nanoparticles, thus leading to a variation in the relative population of interfacial frozen spins. Since the shell thickness remains constant in all of the studied particles, $H_E$ is directly related to the moment of irreversible spins and inversely related to the reversible spins. Also, $H_E$ depends linearly on the ratio of number of frozen spins ($M_f$) to reversible spins. As one can see clearly in Figure 7e, with decrease in the particle size from 15 to 8 nm, $M_f$ first increases, reaches a maximum for the 10 nm particles, and then decreases for the smaller particles. This finding gives a simple explanation for the temperature dependence of $H_E$ and for the relationship between the population of interfacial frozen spins and the strength of exchange coupling between the core and shell. We note that while the $H_E$ of the 15 nm sample is smaller than that of the 8 nm sample, an opposite trend is observed for $M_f$ in these two cases (Figure 7e).

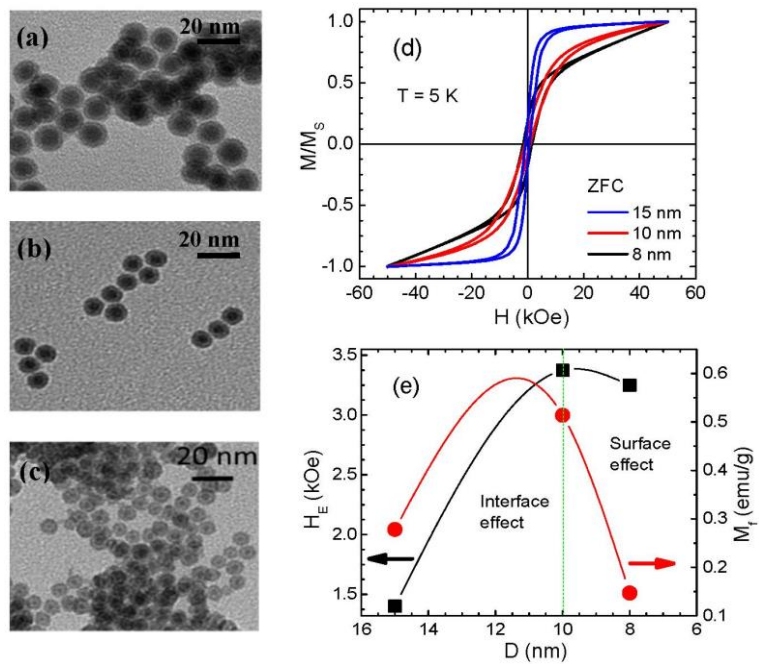

**Figure 7.** TEM images of Fe/$\gamma$-Fe$_2$O$_3$ core/shell nanoparticles with three different sizes of 15 nm (**a**), 10 (**b**) and 8 nm (**c**) while retaining the same shell thickness of 2 nm; (**d**) Normalized magnetization vs.

magnetic field for these three MNP sizes; (e) The EB field ($H_E$) and the net moment of frozen spins ($M_f$) are plotted as functions of particle size [29].

In order to shed light on this, we have quantified the SPM and PM contributions to the magnetization by fitting the room temperature M-H data to the Langevin function with an added linear term [29]. An increasing contribution to the magnetization from the PM susceptibility with the decrease in particle size consistently confirms that surface spins are more disordered and their impacts on the EB are stronger in the case of smaller particles. These results reveal that *there exists a critical particle size (~10 nm), above which the spins at the interface between Fe and γ-Fe₂O₃ contribute primarily to the EB, but below which the surface spin effect is dominant.* Our findings provide physical insights into the collective contributions of interface and surface spins to the EB in core/shell nanoparticle systems, knowledge of which is the key to manipulating EB fields in magnetic nanostructures for spintronics applications.

As we noted above in Section 3, no EB has been observed in spherical Fe₃O₄ MNPs for particle sizes as small as 3 nm [58,61]. The nature of surface spin ordering is also not greatly altered by varying particle shape [54–58]. This gives a natural explanation for the observed small EB effect in hollow Fe₃O₄ MNPs of ~4.5 nm thickness, and that the interface spins play a dominant role in inducing the EB effect in the ~14 nm Fe/Fe₃O₄ core/shell MNPs [30,31]. By contrast, the surface spins are largely disordered in γ-Fe₂O₃ MNPs for particle sizes just below 10 nm (Figure 4c), where the degree of disorder of the surface spins in the outer layer to which the ferromagnetically ordered spins of the core are coupled has been believed to be crucial for achieving EB [28,29]. The surface spins of these MNPs become even more disordered when the spherical symmetry is broken [19]. Therefore, the surface spins of the γ-Fe₂O₃ shell of 2–4 nm thickness are expected to be largely disordered, contributing significantly to the EB (more precisely, the M-H loop shift) in the case of Fe/γ-Fe₂O₃ MNPs with diameters less than 10 nm [28,29]. These observations shed light on the different origins of EB in between the magnetite- and maghemite-based core/shell MNP systems [27–36,73].

## 5. Exchange Bias Effect in Hollow Nanoparticles

Hollow magnetic nanoparticles have been especially attractive for the study of EB due to the formation of an additional inner surface [40,41,44]. The increased surface area can give rise to an enhanced spin disorder and hence a higher anisotropy. During the last few years, there have been several reports on the synthesis of hollow nanoparticles using different methods [84–86]. The formation of hollow nanoparticles has been related to the Kirkendall effect, which is related to the difference in the diffusivities of atoms at the interface of two different materials, causing supersaturation of lattice vacancies. These supersaturated vacancies can condense to form "Kirkendall voids" close to the interface [87,88]. On the nanometer scale the Kirkendall effect can be controlled to fabricate hollow MNPs from initial core/shell MNPs [29,31,89]. The hollow MNPs are often polycrystalline, with multiple crystallographic domains that are randomly oriented with differentiated local anisotropy axes (see inset of Figure 8a for the case of ~9 nm γ-Fe₂O₃ hollow MNPs of ~2 nm shell thickness).

Owing to the hollow morphology that leads to an enhancement in magnetic anisotropy, higher values of the EB field and blocking temperature have been reported for NiFe₂O₄ and CoFe₂O₄ hollow nanoparticles [89,90] than their solid counterparts. While Ong et al. [31] have reported a small FC hysteresis loop shift for the Fe₃O₄ hollow MNPs (diameter of ~14 nm, shell thickness of ~4.5 nm), Cabot et al. [40] have observed a very strong shift of the M-H loop (over 3 kOe) for the γ-Fe₂O₃ hollow MNPs (diameter of ~8.1 nm, shell thickness of ~1.6 nm) when cooling the particles in a field of 5 T. In the latter case, the observed large loop shift has been attributed to the "minor loop" phenomenon (Figure 8a) [40], rather than an intrinsic EB effect [1]. Interestingly, a Monte Carlo simulation study has predicted that the crystallographic anisotropy would dominate the low-temperature magnetic behavior in γ-Fe₂O₃ hollow MNPs if the number of surface spins increases [40]. In addition, it has been experimentally noted that the degree of disorder of spins located near the outer and inner layers may differ significantly in hollow magnetic MNPs upon size reduction, while the shell thickness of the particles remains the same [89,90]. These observations have led us to two fundamental and important questions: *Can one tune "minor loop" to "exchange bias" effect in hollow magnetic nanoparticles by varying the number of surface spins?* and *Can one decouple the contributions of the inner and outer surface spins to the EB in a hollow nanoparticle system?*

To address these questions, we have performed a comparative study of the magnetic properties including EB in monodisperse hollow $\gamma$-Fe$_2$O$_3$ MNPs with two distinctly different sizes of ~9 nm and ~18 nm [41]. More details about the synthesis of these hollow nanoparticles can be found in Appendix B. HRTEM images confirmed shell thicknesses of 2 nm and 4.5 nm for the 9 nm and 18 nm nanoparticles, respectively. We have observed anomalously large horizontal shifts and open hysteresis loops in fields as high as 9 T for the 9 nm MNPs, corresponding to a "*minor loop*" of the hysteresis loop as reported previously by Cabot et al. [40] for 8 nm $\gamma$-Fe$_2$O$_3$ hollow MNPs. Meanwhile, the loop shift observed for the 18 nm MNPs manifests an *intrinsic* EB effect. Relative to the 18 nm solid MNPs, a much stronger EB effect has been observed in the 18 nm hollow MNPs, demonstrating the important role of inner surface spins in enhancing the EB effect in $\gamma$-Fe$_2$O$_3$ hollow MNPs. We note herein that the Monte Carlo simulation study shows that the magnetic exchange interactions between spins with different crystallographic easy axes inside the shell have a noticeable, but not dominant, influence on the hysteresis loops [40]. A quantitative analysis of the SPM and PM contributions to the magnetization has also been performed on these MNPs by fitting the room temperature M-H data to the Langevin function with an added linear term using the following expression:

$$M(H) = M_s^{SPM}\left[\coth\left(\frac{\mu H}{KT}\right) - \left(\frac{\mu H}{KT}\right)^{-1}\right] + C^{PM}H, \qquad (3)$$

where $M_s^{SPM}$ is the saturation magnetization of the SPM part and $\mu$ is the average magnetic moment of SPM particles. $C^{PM}$ is the susceptibility of the paramagnetic contribution that is linear with the magnetic field, $H$.

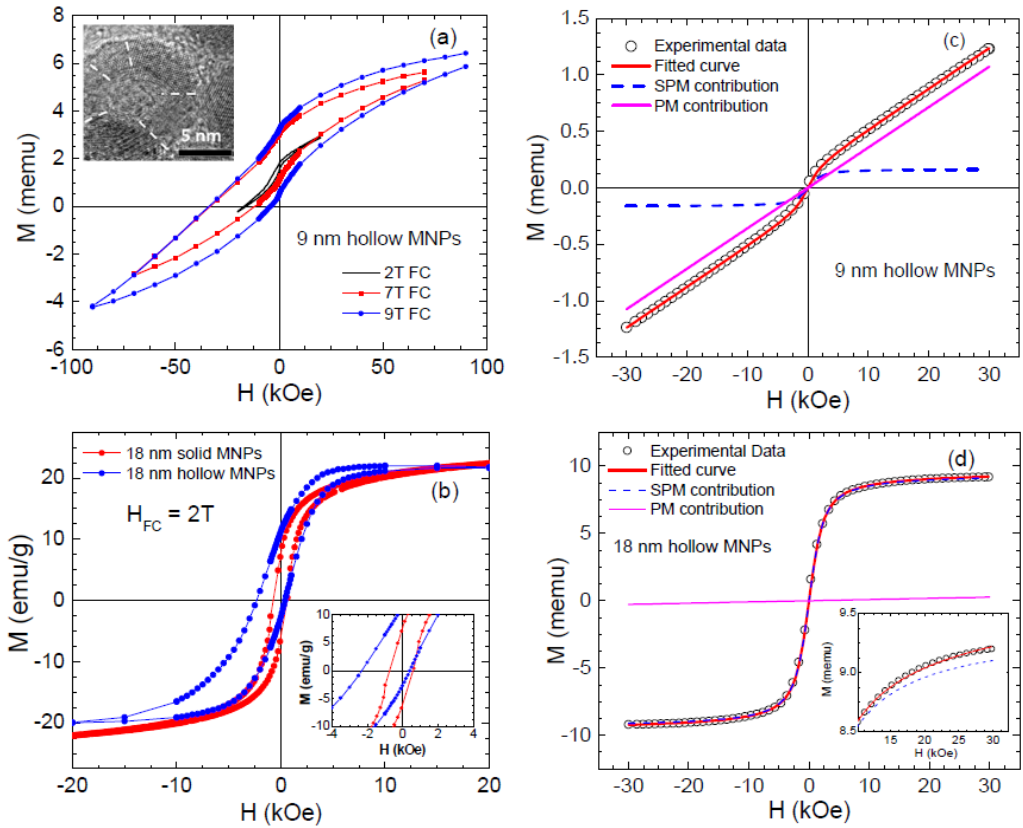

**Figure 8.** The FC M-H loops taken at 5 K for (**a**) 9 nm hollow nanoparticles for different cooling fields of 2 T, 7 T, and 9 T; and (**b**) 18 nm solid and hollow nanoparticles for a cooling field of 2 T; An enlarged portion of the FC M-H curves is shown in the inset of Figure 8b. The M-H curves at 300 K are fitted (red) to Equation (3); The blue and magenta (dashed) curves represent the simulated SPM and PM contributions extracted from the experimental data using the fitting parameters for the (**c**) 9 nm and (**d**) 18 nm hollow nanoparticles [41].

As one can see clearly in Figure 8c,d, SPM susceptibility contributes only 13% to the total magnetic moment for the case of 9 nm hollow MNPs, while the rest of it (87%) comes from the PM susceptibility. By contrast, SPM susceptibility contributes 97% to the total magnetic moment and only a 3% contribution comes from the PM susceptibility for the case of the 18 nm hollow MNPs. Since a highly linear contribution to the magnetization results mainly from the uncompensated spins at the shell surfaces, these results reveal a larger number of disordered surface spins present in the 9 nm hollow particles than in the 18 nm hollow particles [41], thus explaining the non-saturation feature of magnetization and the smaller value of magnetization for the 9 nm hollow MNPs (Figure 8c). This can be reconciled with our recent study that shows that the magnetic relaxation in the $\gamma$-Fe$_2$O$_3$ hollow nanoparticle ensembles is best described by a non-interacting particle model, as the dominant role of disordered surface spins and severely reduced particle magnetization renders the influence of dipolar interactions negligible in determining the low-temperature magnetic behavior [42,43].

On the other hand, it has been noted that the presence of voids at the core/shell interface influences the coupling between interface spins and hence the magnetic properties [31,89,90]. Since the $\gamma$-Fe$_2$O$_3$ hollow MNPs have been produced directly by further oxidizing the Fe/$\gamma$-Fe$_2$O$_3$ core/shell MNPs, which became hollow via the nanoscale Kirkendall effect [31,41,42], it is essential to understand *how the magnetic properties of a core/shell nanoparticle system are modified when the core/shell morphology transforms into the core/void/shell and the hollow structure*. This has recently motivated us to study the evolution of the structural and magnetic properties of Fe/$\gamma$-Fe$_2$O$_3$ core/shell, core/void/shell, and hollow MNPs with two different sizes of 8 nm and 12 nm [43]. We find that as the nanoparticles become hollow, both their shell thickness and total diameter increase. As the morphology changes from core/shell to core/void/shell, the magnetization of the system decays and inter-particle interactions become weaker, while the effective anisotropy and the EB effect increase. The changes are more drastic when the MNPs become completely hollow. Noticeably, the morphological change from core/shell to hollow increases the mean blocking temperature for the 12 nm particles but decreases for the 8 nm particles. While the low-temperature magnetic behavior of the 12 nm MNPs changes from a collective SSG system mediated by dipolar interactions for the core/shell MNPs to a frustrated cluster glass-like state for the shell nanograins in the hollow morphology, the magnetic behavior is more similar for the 8 nm core/shell and hollow particles, and a conventional spin glass-like transition is obtained at low temperatures. In case of the hollow MNPs, the coupling between the inner and outer spin layers in the shell gives rise to an enhanced EB effect, which increases with increasing shell thickness. All of these findings point to the importance of inner and outer surface spin disorder giving rise to surface anisotropy and EB, and reveal a new path toward tuning EB fields in hollow MNP systems by varying the number of surface spins (the thickness of the shell and/or the diameter of MNPs).

In an attempt to decouple relative contributions of the inner and outer surface spins to the EB in a hollow nanoparticles system, we have performed Monte Carlo simulations of the hysteresis loops of a hollow $\gamma$-Fe$_2$O$_3$ particle with a diameter of 15 nm [44]. The simulations have been performed at the atomistic level considering classical spins placed at the nodes of the spinel lattice of $\gamma$-Fe$_2$O$_3$, as explained in Appendix A and in Reference [40].

By using this model, we have been able to simulate the alignment of spins in the inner and outer surface layers of the shell under a dc field, revealing clear differences in the response of the magnetic ions depending on the surface (see Figure 9a–c). Figure 9d–f presents a series of simulated M-H loops, with a field step, $dh$ = 5, for a particle with surface Néel anisotropy with constant, $ks$ = 30 K and uniaxial anisotropy for core spins, $kc$ = 0.01 K (equal to the bulk value of $\gamma$-Fe$_2$O$_3$) calculated at low temperature ($T$ = 0.1 K). As observed, the global shape of the M-H loops of surface spins is qualitatively similar in both cases. However, the reversal of the interior spins is clearly influenced by the presence of additional surface spins at the inner surface of the hollow MNPs. An increased shift in the core M-H loop is observed for the hollow MNPs. The more elongated shape together with reduced remanence and increased coercive field of the simulated loops are a consequence of the enhanced disorder due to the larger number of surface spins, with those spins at the outer surface layer exhibiting a higher degree of frustration. If we compare both solid and hollow morphologies, we can observe that while the coercive field and loop shifts for the surface contributions are almost identical (Figure 9e), the core magnetization contributions are different (Figure 9f). This, once again, suggests that the additional inner surface spins in the hollow MNPs are responsible for the differences in the magnetic behavior with respect to their solid counterparts. Finally, we must remark the presence of a jump in the magnetization in the simulated M-H loops around zero field values in the hollow MNPs, which has also been observed experimentally [41–44]. This feature is not visible for the solid MNPs, so one may attribute it to the effect of inner surface spins.

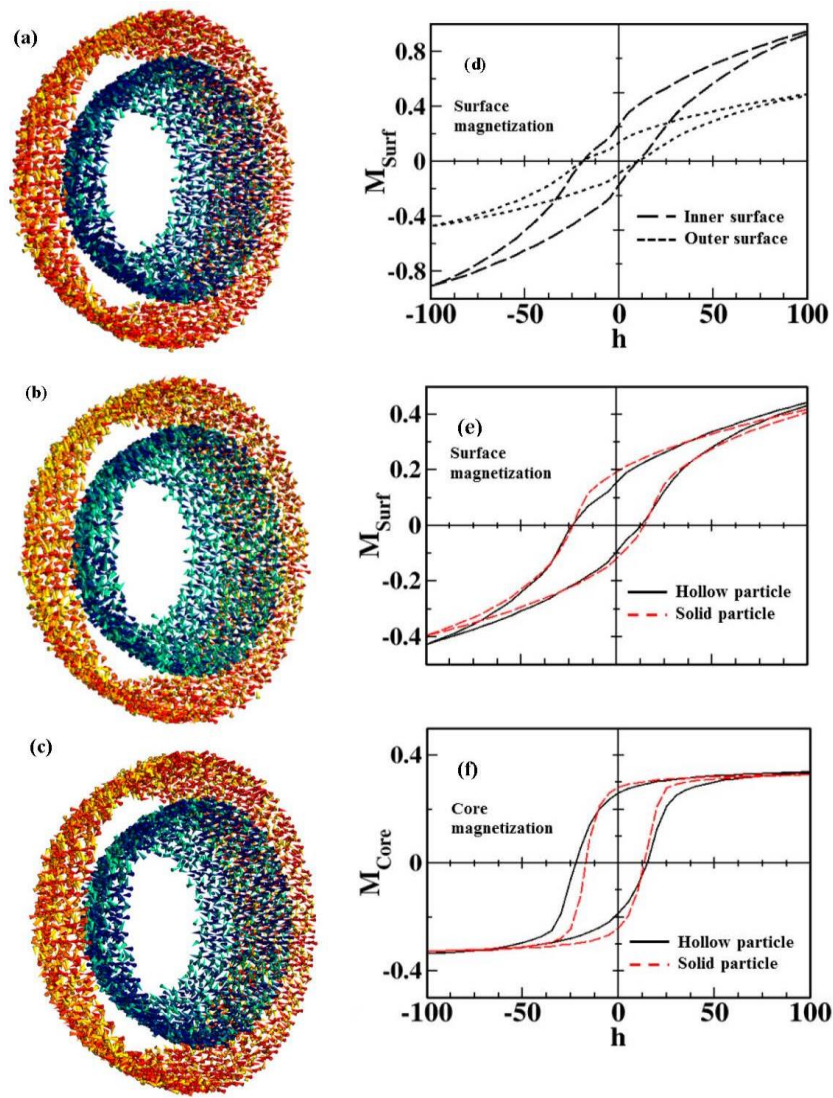

**Figure 9.** Snapshots of the outer and inner surface spin configurations subject to varying magnetic fields (**a**) $h$ = 100 (the maximum positive applied field), (**b**) $h$ = 0 (remanence at the upper branch), and (**c**) $h$ = −100 (the maximum negative applied field). Spins have been colored with a gradient from dark-red/dark-blue (outer/inner surface) for spins along the field direction to yellow/green (outer/inner surface) for spins transverse to the field direction. Only a slice of the spin configurations of a hollow

nanoparticle close to the central plane and perpendicular to the field direction is shown; Low temperature ($T = 0.1$ K) simulated hysteresis loops for a hollow particle with the same dimensions as in the experiments (shell thickness of 2.5 nm) and surface and core anisotropy constants $k_s = 30$ K and $k_c = 0.01$ K. (**d**) Contributions of the spins at the inner (long dashed lines) and outer (short dashed lines) surfaces of the hollow particle to the total hysteresis loop with 1000 Monte Carlo steps. (**e**) Contribution of all surface spins to the total magnetization for the hollow MNP (continuous black line) is compared to that of a solid MNP with the same diameter (dashed red line). (**f**) The same as in panel (**e**) but for the interior spins contribution [44].

## 6. Exchange Bias Effect in Hybrid Composite Nanoparticles

An issue of potential interest is if the surface spin alignment in nanoparticles could be influenced by forming interfaces with other non-magnetic materials [37,91–96]. If this is indeed possible, then it would provide an excellent mechanism to control the exchange coupling between the surface and core spins in individual MNPs leading to novel magnetic properties. Desautels et al. [91,92] have reported that coating ~7 nm $\gamma$-Fe$_2$O$_3$ MNPs with a metallic Cu shell of 0.5 nm thickness significantly decreases the intrinsic surface spin disorder and hence the EB field with respect to the uncoated $\gamma$-Fe$_2$O$_3$ MNPs. Interestingly, they have shown that there exists an interfacial monolayer of CuO in the Cu-coated $\gamma$-Fe$_2$O$_3$ MNPs and this layer actually stabilizes the disordered surface spins of the $\gamma$-Fe$_2$O$_3$ MNPs [92]. In another case, Shevchenko et al. [93] have reported the observation of a strong EB effect in Au/$\gamma$-Fe$_2$O$_3$ core/shell MNPs. However, no comparison to the bare $\gamma$-Fe$_2$O$_3$ hollow MNPs was performed, leaving an unanswered question about the influence of Au on the magnetic properties, including EB, in these nanostructures. Recently, Pineider et al. [94] have revealed that a magnetic moment is induced in the Au domain when the Fe$_3$O$_4$ shell contains a reduced iron oxide phase (FeO) in direct contact with a noble metal (Au). This is in contrast with the case of Au/$\gamma$-Fe$_2$O$_3$ nanocrystals, in which $\gamma$-Fe$_2$O$_3$ has been shown to disfavor such a spin polarization transfer process [94]. A later study of Feygenson et al. [95] has indeed confirmed that the charge transfer from the Au nanoparticles is responsible for a partial reduction of the Fe$_3$O$_4$ into the FeO phase at the interface with Au nanoparticles, and that the coupling between the Fe$_3$O$_4$ and FeO induces an EB effect. Since FeO is formed at the interface between Fe$_3$O$_4$ and Au [94], in the case of Au/$\gamma$-Fe$_2$O$_3$ core/shell MNPs [93] the absence of FeO suggests that the EB effect is attributed to the presence of $\gamma$-Fe$_2$O$_3$ hollow morphology. Figure 10a,b shows TEM images of two types of MNPs with different interfaces [94], through which the different EB mechanisms, as recently proposed for these nanostructures and others [93,95], can be understood.

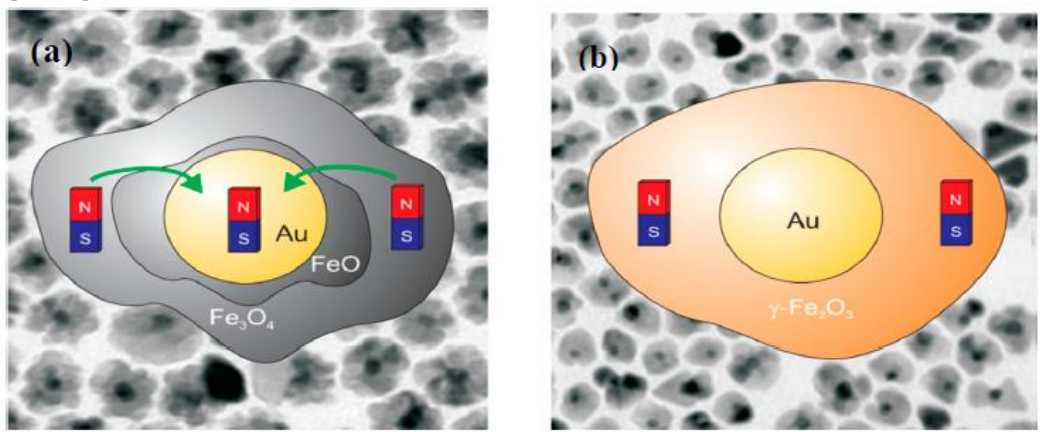

**Figure 10.** TEM images of (**a**) Au/FeO/Fe$_3$O$_4$ and (**b**) Au/Fe$_2$O$_3$ nanostructures; Reprinted with permission from [94].

It has been noted that the magnetic properties, including EB, in the Au-Fe$_3$O$_4$ nanostructures depend sensitively on their sample preparation methods and measurements [37,94,96–99]. Recent advances in chemical synthesis [97] have given us the opportunity to investigate the novel magnetic properties of an Au-Fe₃O₄ composite nanoparticle system [37], where one or more Fe₃O₄ MNPs (mean size, ~9 nm) are attached to an Au seed particle (mean size, ~8 nm) forming "dumbbell"- and "flower"-like morphologies (Figure 11a and its inset). We have observed a noticeable EB in Au-Fe₃O₄ MNPs, with the flowers showing much stronger EB in comparison with the dumbbells, while the bare 9 nm Fe₃O₄ MNPs show no EB effect (see inset of Figure 11b). It has been theoretically predicted that in most single-phase nanoparticle systems, the spherical symmetry of the disordered spins leaves no net surface anisotropy and EB should not be observed [100]. This prediction explains the absence of EB in our 9 nm spherical Fe₃O₄ MNPs and is in full agreement with the previous experimental observations of no EB in spherical Fe₃O₄ MNPs, even down to 5 nm [54]. From these analyses, it is natural to hypothesize that while the disordered spins in the surface layer of the 9 nm Fe₃O₄ nanoparticles is insufficient to couple with the magnetically ordered core to induce an EB, the presence of Au in the Au-Fe₃O₄ MNPs could break the spherical symmetry of the Fe₃O₄ MNPs leaving a shape asymmetry, which enhances the disordering of spins in the surface layer and consequently results in EB in these systems.

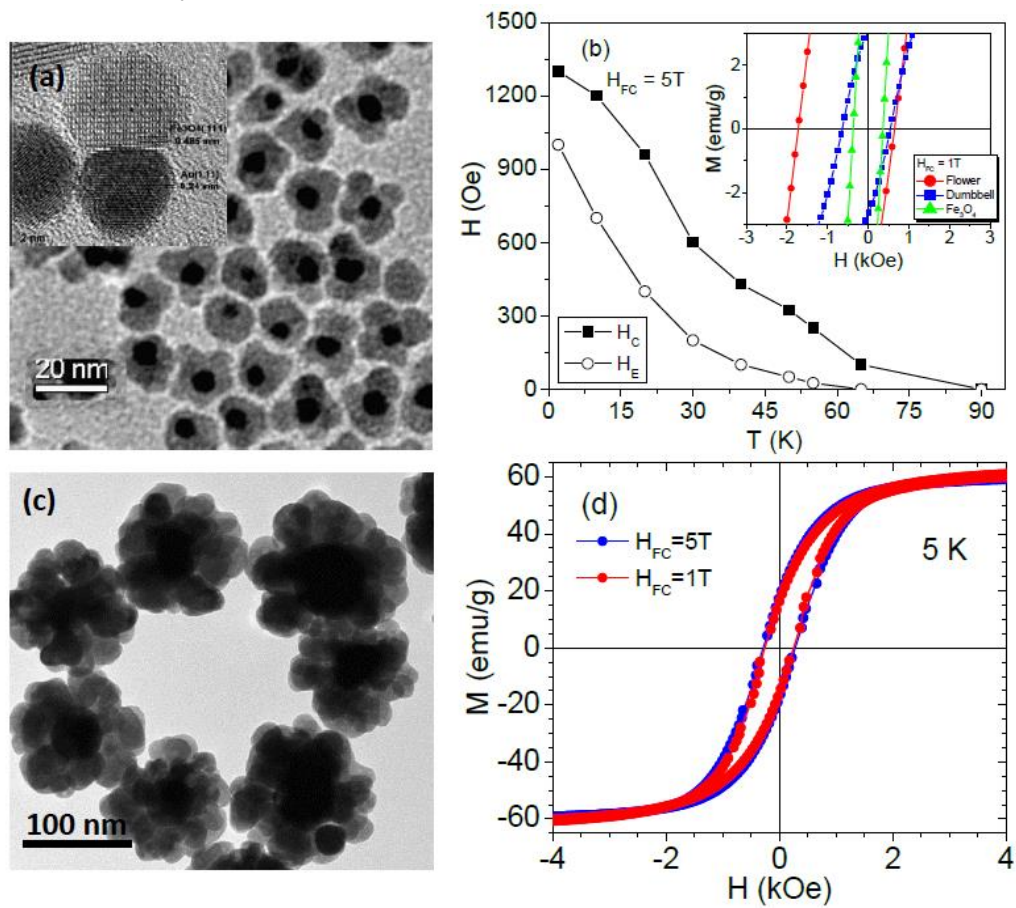

**Figure 11.** (**a**) TEM image of flower-like Au-Fe₃O₄ nanoparticles. Inset shows an HRTEM image of dumbbell-like Au-Fe₃O₄ nanoparticles; (**b**) Temperature dependence of exchange bias field ($H_E$) for the flower-like Au-Fe₃O₄ nanoparticles and its inset shows the FM hystereses loops for the bare Fe₃O₄, dumbbell- and flower-like Au-Fe₃O₄ nanoparticles [37]. (**c**) TEM image of flower-like Ag-Fe₃O₄ nanoparticles (the diameter of Ag is ~45 nm and the diameter of Fe₃O₄ is ~20 nm) and (**d**) the FC hysteresis loops of these nanoparticles.

To test this hypothesis, we have performed magnetic measurements on (i) Au-Fe₃O₄ dumbbells having diameters of 4 nm or 6 nm Au and 18 nm Fe₃O₄ (where the 18 nm Fe₃O₄ nanoparticles should exhibit no EB regardless of the shape) and (ii) 8 nm Au–9nm Fe₃O₄ dumbbells after the selective chemical etching of the Au phase using a potassium triiodide (KI/I₂) etchant. We have indeed observed the EB in both dumbbell samples with Fe₃O₄ as large as 18 nm, as well as in the 8 nm Au–9 nm $Fe_3O_4$ dumbbells after the selective chemical etching of Au. These results consistently rule out surface spin disorder and shape asymmetry as the driving force for the EB behavior observed in the Au-$Fe_3O_4$ MNPs. From the synthesis perspective, it has been noted that the synthesis of the Au-$Fe_3O_4$ MNPs relies on a Fe (0) precursor, not an iron ion [97]. As a result, the deposition of iron atoms on the surface of the Au seed particle can be compared to the atomic layer deposition of Fe on an Au substrate, a system for which it has been well-documented that diffusion can easily occur [101]. In other words, Fe could diffuse into the Au without forming islands, which, in effect, resulted in an AuFe spin glass. However, X-ray magnetic circular dichroism (XMCD) studies seem to rule out this hypothesis, which is in agreement with the previous observation of Pineider et al. [94]. On the other hand, the development of stress at the heterogeneous interface between $Fe_3O_4$ and Au has been reported for the Au-$Fe_3O_4$ dumbbells [102]. The mechanical modeling analysis revealed that the development of stress occurred due to different thermal expansion coefficients of Au and $Fe_3O_4$ at the interface of Au-$Fe_3O_4$, and was on the order of 1–5 GPa. It was also shown that compacting 20 nm $Fe_3O_4$ nanoparticles under external pressure (1–5 GPa) resulted in the development of surface spin disorder and hence the EB [63]. We note that the magnitude of external stress applied to the 20 nm $Fe_3O_4$ MNPs is of the same order as that generated across the Au-$Fe_3O_4$ interface of the dumbbells. As a consequence of the interfacial stress, the $Fe_3O_4$ MNPs in the dumbbells could develop surface spin disorder by way of energy minimization. The disordered surface spins are highly anisotropic, which is consistent with the rise in their effective magnetic anisotropy. The disordered surface spins undergo exchange coupling with the core moments, resulting in the EB effect in both dumbbells and flowers. Relative to the dumbbells, the larger EB effect observed for the flowers seems to suggest that a larger EB field is observed when there is a larger number of interfaces. However, our recent study of the magnetic properties of flower-like Ag-$Fe_3O_4$ MNPs (see more information on their synthesis on Appendix B, and a TEM image of these MNPs in Figure 11c; the diameters are ~45 nm and ~20 nm for Ag and $Fe_3O_4$, respectively) has revealed that if the size of $Fe_3O_4$ MNPs is large (>>10 nm, for those sizes the effect of surface spin disorder is small [54,58]), the interfacial stress effect induced by Ag on the ordering of its surface spins is small which is therefore insufficient to induce an EB (Figure 11d). So, it will be interesting to tune the particle size of $Fe_3O_4$ to find an optimal size for which the largest EB effect can be achieved in this type of hybrid nanostructure.

To complement our experimental findings of the EB effects for the dumbbell- and flower-like Au-$Fe_3O_4$ MNPs (Figure 11b), we have performed Monte Carlo simulations of an atomistic model of $Fe_3O_4$ for Heisenberg spins. Details of the simulation can be found in Appendix A. Values for the $J_{ij}$ between spins with tetrahedral and octahedral coordination have been taken from the available literature [72,103]. Fe ions with reduced coordination with respect to bulk are considered to be surface spins with Néel type anisotropy and anisotropy constant $k_S$, while core spins have uniaxial anisotropy along the field direction with anisotropy constant $k_C$. This time, the value of the anisotropy constants, expressed in units of K/spin have been taken as $k_C = 0.01$ and $k_S$ varying in the range 0.01–30. In order to model the geometry of the dumbbell particle, a sphere of radius 5.5a (being 'a' the unit cell size) truncated by a sharp facet where magnetite contacts Au has been considered. On the other hand, for the flower arrangement, four overlapping spheres of radius 5a surrounding a spherical hole that stands for the central Au component have been employed.

In Figure 12, we present the simulated hysteresis loops after an FC process at $h_{FC} = 100$ K for the spherical-, flower-, and dumbbell-shaped nanoparticles. In order to recreate the effect of interfacial stress and consequential disorder in the surface spins, we have assigned increased Néel surface anisotropy for the surface spins ($k_S = 30$) as compared to the core spins which are assumed to have the same anisotropy as the bulk ($k_C = 0.01$). By considering an increased surface anisotropy ($k_S = 30$, red circles in Figure 12a,b) a slower approach to saturation and high field irreversibility is obtained, along with the expected observation of EB, which is practically absent when we consider that the surface and core anisotropies are the same, $k_S = k_C$ (blue squares in Figure 12a,b). Moreover, it can be observed that the horizontal shift of the loops is noticeably higher for the flower than for the dumbbell arrangement, as also observed experimentally [37], which demonstrates that the EB can be tuned by the increase of the contact interfaces between Au and the MNPs. The case for the dumbbell-shaped nanoparticles can be reproduced by assigning the same anisotropy values to the surface and core moments. We observe that for $Fe_3O_4$ in both geometries, when the surface anisotropy is equal to the core one ($k_S = k_C = 0.01$), neither loops exhibit horizontal shifts after an FC (blue squares in Figure 12a,b). We note that for lower anisotropy values ($k_S$= 0.01), the hysteresis loops have a more squared shape than those measured for $k_S$ = 30, which indicates high field linear susceptibility, irreversibility and lack of saturation both for dumb-bell and flower arrangements, as also observed experimentally. This last result can be attributed to the contribution of surface spins (green squares in the inset of Figure 12b) that exhibits a hysteresis loop typical of a frustrated material and dominates the magnetization reversal of the whole particle. However, the reversal shown by core spins (yellow circles in the inset of Figure 12b), is more coherent despite the influence of the surface spins. The higher value of the remanent magnetization for the dumbbell than for the flowers is also in agreement with the experimental results. The high degree of disorder at the particle surface is also corroborated by the snapshots of the spin configurations presented in Figure 12c–f, where one can notice that, even after the high FC process (Figure 6c,e), only the core spins (drawn in lighter colors) are aligned along the anisotropy axis while surface spins remain highly disordered even at low temperatures. Snapshots taken at field values close to the coercive field (Figure 12d,f) show a more coherent reversal of core spins that are dragged by the disordered shell of surface spins (drawn in darker tones).

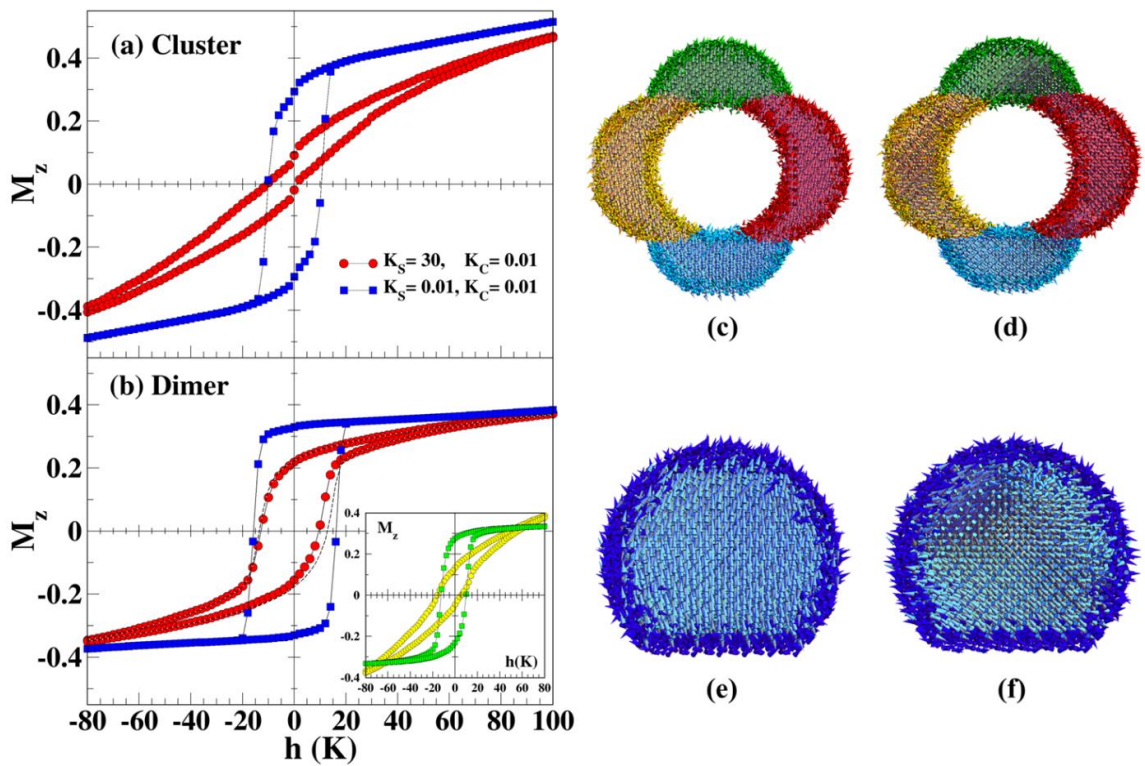

**Figure 12.** Low temperature hysteresis loops simulated after cooling in a magnetic field $h_{FC}$ = 100 K as computed by Monte Carlo simulations of individual nanoparticles with cluster (**a**) and dimer (**b**) geometries. The non-magnetic metal is simulated as a hole in the middle for the cluster geometry and a sharp facet for the cluster. Panels (**a,b**) show hysteresis loops of a particle with cluster and dimer geometry, respectively, for two different values of the surface anisotropy constant: $k_S$= 0.01 (blue squares) equal to the core value $k_C$= 0.01, and increased surface anisotropy $k_S$= 30 (red circles). The dashed lines in (**b**) stand for a spherical particle of the same size as the dimer. The Inset shows the contribution of the surface (yellow circles) and core (green squares) spins of a dimer particle to the hysteresis loop for $k_S$ = 30. Snapshots of the spin configurations for cluster (**c,d**) and dimer (**e,f**) particles for $k_S$= 30 (red circles) obtained at the end of the FC process (**c,e**) and at the coercive field point of the decreasing field branch (**d,f**) of the hysteresis loops. For clarity, only a slice of width 4a along the applied field direction and through the central plane of the particles is shown. Surface spins have darker colors and core spins have been colored lighter; Reprinted with permission from [37].

## 7. Prospective Applications of Exchange-Coupled Nanoparticles

It has been reported that the exploitation of EB provides a novel approach to overcoming the superparamagnetic limit and increasing the thermoremanence of magnetic nanoparticles for use in advanced disk media and spintronic devices [3,4]. Therefore, we summarize in Table 1 the maximum EB fields and other important parameters for selected iron oxide-based nanoparticle systems [19,27,31–35,37,41,58,65,73,92,93,104–107]. It can be observed that while no EB effect has been reported for spherical $Fe_3O_4$ MNPs with diameters as small as 3 nm [58], a remarkable EB effect ($H_E$ = 133 Oe) has been obtained for hollow $Fe_3O_4$ MNPs (the outer diameter is 16 and the shell thickness is 4.5 nm) [31]. The EB effect is greatly enhanced in the case of hollow $\gamma$-$Fe_2O_3$ MNPs relative to their solid counterparts (see Table 1). These observations highlight a significant effect of hollow morphology on the ordering of surface spins and hence the coupling between the magnetically ordered core and the magnetically disordered shell, resulting in the EB. As compared to the bare $Fe_3O_4$ or $\gamma$-$Fe_2O_3$ MNPs, the EB effects are largely enhanced when these MNPs are coupled with other magnetic materials such as Fe and FeO. It appears that the AFM/FI $FeO/Fe_3O_4$ core/shell MNPs [31] yield larger values of $H_E$ as compared to the FM/FI $Fe/Fe_3O_4$ core/shell MNPs of similar size [32]. Interestingly, Wetterskog et al. reported a large value of $H_E$ (~1550 Oe) for the $FeO/Fe_3O_4$ core/shell nanocubes and showed how the structure was transformed into a single phase material upon topotaxial oxidation [106]. Among the reported EB materials, the FM/FI $Fe/\gamma$-$Fe_2O_3$ core/shell MNPs of 11 nm diameter show the largest value of $H_E$ (~6300 Oe). As we discussed above in Section 2, the magnetic properties of iron oxide MNPs depend sensitively on their size, shape, and morphology. Since the inter- and intra-particle interactions may differ significantly in between diluted and condensed MNP systems, their impacts on the magnetic behavior and EB may be different. Therefore, one should be careful when comparing the EB fields and other magnetic parameters among samples prepared under different synthesis conditions [33,105].

In addition to its potential spintronics applications [108,109], EB is an important tool in understanding fundamental nanoscale spin ordering, knowledge of which is essential to tailor the anisotropic magnetic properties of nanoparticle systems for potential rare-earth-free permanent magnets applications [110] and for biomedical applications such as magnetic hyperthermia [42,111–113] via a controllable exchange coupling mechanism. In particular, Lee et al. have shown that the effective anisotropy can be tuned by using exchange-coupled core/shell nanoparticles (exchange anisotropy) which, in effect, increases the heating efficiency of MNPs [111]. In case of the exchange-coupled $FeO/Fe_3O_4$ nanoparticles, Khurshid et al. have compared two sets of comparable particles: The spheres with 1.5 times bigger saturation magnetization than the cubes, and the cubes with 1.5 times bigger effective anisotropy than the spheres, while keeping the other parameters the same. The authors have proved that increasing the effective anisotropy of the nanoparticles gives rise to a greater heating efficiency than increasing their saturation magnetization. The improved SAR values have also recently been reported in $Fe_{1-x}O/Fe_{3-\delta}O_4$ core/shell nanocubes, resulting from a gradual transformation of the $Fe_{1-x}O$ core to $Fe_{3-\delta}O_4$ [112]. In both cases, however, the obtained SAR values are relatively small especially at AC magnetic fields below 400 Oe, rendering these nanomaterials unsuitable for active magnetic hyperthermia-based cancer therapy. For the $Fe/\gamma$-$Fe_2O_3$ core/shell nanoparticles, a modest heating efficiency has recently been reported, resulting mainly from the strong reduction in saturation magnetization caused by the shrinkage of the core with time [113]. However, for sizes above 14 nm, the shrinkage process is much slower and the obtained heating efficiency is better than the one exhibited by conventional solid nanoparticles of the same size. Despite these recent efforts, there are still several questions about the efficiency of tuning magnetic anisotropy of the nanoparticles to achieve large SAR at low AC magnetic fields.

**Table 1.** Measured maximum exchange bias field ($H_E$), types of magnetic coupling, mean blocking temperature ($T_B$), and diameter and shell thickness of iron oxide-based nanoparticles.

| Material | Total Diameter, $d$ (nm) | Types of Magnetic Coupling | Loop Shift, $H_E$ (Oe) | Blocking Temperature, $T_B$ (K) | Reference |
|---|---|---|---|---|---|
| $Fe_3O_4$ (solid) | 9 | FI | 0 | 36 | [37] |
| $Fe_3O_4$ (solid), under 5 GPa pressure | 20 | FI | 800 | 40 | [63] |
| $Fe_3O_4$ (hollow) | 16/4.5 | FI | 133 | -- | [31] |
| $Fe/Fe_3O_4$ (C/S) | 14/2.5 | FM/FI | 1190 | 115 | [31] |
| $FeO/Fe_3O_4$ (C/S) | 14/3.5 | AFM/FI | 471 | -- | [32] |
| $FeO/Fe_3O_4$ (C/S) | 35/4 | AFM/FI | 2250 | 270 | [33] |
| $FeO/Fe_3O_4$ (C/S) | 18/6 | AFM/FI | 1000 | 210 | [105] |
| $FeO/Fe_3O_4$ (C/S) | 24/5 | AFM/FI | 3700 | 260 | [105] |
| $FeO/Fe_3O_4$ (C/S) | 10/0.6 | AFM/FI | 1700 | 220 | [34] |
| $FeO/Fe_3O_4$ (C/S) | 20 | AFM/FI | 1550 | >275 | [106] |
| $Fe_3O_4/FeO$ (C/S) | 6 | FI/AFM | 514 | -- | [104] |
| $Fe_3O_4/\gamma\text{-}Fe_2O_3$ (C/S) | 12 | FI/FI | 140 | 105 | [35] |
| $Au/Fe_3O_4$ (dumbbell) | 8/9 | FI | 260 | 65 | [37] |
| $Au/Fe_3O_4$ (flower) | 8/9 | FI | 500 | 90 | [37] |
| $Au/FeO/Fe_3O_4$ (C/S) | 10/14 | AFM/FI | 860 | 250 | [92] |
| $\gamma\text{-}Fe_2O_3$ (solid) | 4 | FI | 750 | 7 | [58] |
| $\gamma\text{-}Fe_2O_3$ (solid) | 7 | FI | 60 | 20 | [65] |
| $\gamma\text{-}Fe_2O_3$ (solid) | $10^a$ | FI | 1500 | 42 | [19] |
| $\gamma\text{-}Fe_2O_3$ (hollow) | 19/4.5 | FI | 960 | 150 | [41] |
| $\gamma\text{-}Fe_2O_3/CuO/Cu$ (C/S) | 7/0.5 | FI/AFM | 18 | 50 | [92] |
| $Fe/\gamma\text{-}Fe_2O_3$ (C/S) | 11 | FM/FI | 6300 | 110 | [27] |
| $\gamma\text{-}Fe_2O_3/CoO$ (C/S) | 5 | FI/AFM | 1200 | -- | [36] |
| $CoO/\gamma\text{-}Fe_2O_3$ (C/S) | 5 | AFM/FI | 2000 | -- | [36] |
| $MnFe_2O_4/\gamma\text{-}Fe_2O_3$ (C/S) | 3.3/0.4 | FI/FI | 86 | -- | [73] |
| $CoFe_2O_4/\gamma\text{-}Fe_2O_3$ (C/S) | 3.1/0.5 | FI/FI | 125 | -- | [73] |
| $Au/\gamma\text{-}Fe_2O_3$ (C/S) | 13/3 | FI | 1200 | 65 | [93] |
| $Fe/\alpha\text{-}Fe_2O_3$ (C/S) | 70/10 | FM/AFM | 78 | -- | [107] |

$^a$ The aspect ratio is 4.

## 8. Concluding Remarks and Outlook

We have shown that the iron/iron oxide core/shell and hollow iron oxide MNPs are excellent model systems for probing the effects of interface and surface spins on nanomagnetism, including EB and related effects, in exchange-coupled magnetic nanostructures. The following findings are worthy of note when working with iron oxide-based nanoparticle systems:

(i)  In case of solid, spherical, single-component MNPs, size reduction to the nanoscale (below 20 nm) has a greater influence on the magnetic ordering of surface spins, namely, a higher degree of disordering of surface spins of $\gamma\text{-}Fe_2O_3$ MNPs as compared to $Fe_3O_4$ MNPs of similar size. As a result, the $\gamma\text{-}Fe_2O_3$ MNPs possess a larger drop in $M_S$ and a greater EB effect.

(ii)  In the case of core/shell MNPs, the interface spins play a more dominant role in triggering the EB effect in $Fe_3O_4$-based core/shell MNP systems such as $Fe/Fe_3O_4$ and $FeO/Fe_3O_4$ MNPs. The surface spins in the $Fe_3O_4$ shell layer have a much lesser contribution to the EB. This is evident from the fact that the EB field is drastically reduced when the Fe core is removed from the core/shell structure. On the other hand, for the case of $\gamma\text{-}Fe_2O_3$-based core/shell nanoparticle systems like $Fe/\gamma\text{-}Fe_2O_3$ MNPs there exists a critical particle size ($d_0 \sim 10$ nm for $Fe/\gamma\text{-}Fe_2O_3$ MNPs of ~2 nm shell thickness), above which the interface spins contribute primarily to the EB, but below which the surface spin effect is dominant. Different values of $d_0$ may be obtained when shell thickness is varied.

(iii) In case of hollow MNPs, both the inner and outer surface spins play important roles in determining the magnetic behavior and EB. The magnetic behavior is not identical at the inner and outer surfaces, with spins at the outer surface layer exhibiting a higher degree of frustration and thus contributing more dominantly to the EB. It is therefore possible to optimize EB by tuning the number of surface spins through varying the thickness of the shell layer and/or the outer diameter of the MNPs. The surface contributions to the coercivity and EB are almost identical for both solid and hollow morphologies, but the core magnetization contributions are different.

(iv) In the case of hybrid composite nanoparticles, coating iron oxide MNPs with other non-magnetic metals can result in different consequences, depending upon the electronic states of metals used and sample synthesis conditions. While the unexpected formation of an intermediate phase during sample synthesis, such as CuO in the case of $\gamma$-Fe$_2$O$_3$/Cu MNPs [92] and FeO in the case of Au/Fe$_3$O$_4$ MNPs [94], complicates the magnetic picture of the systems, it opens up a new pathway for tailoring the anisotropic magnetic properties of these nanostructures if this intermediate phase can be controllably engineered.

Furthermore, we note that while the magnetic ordering of surface spins can be somehow simulated, the exact configurations of interface spins in both bilayer FM/AFM thin film and core/shell nanoparticle systems are still unknown. *How interface spins are formed and oriented in a flat-type interface for the case of a bilayer thin film and in a curve-type interface for the case of a core/shell MNP system represents one of the most challenging tasks for our current understanding.* In this context, the observation of EB in a FM/SG thin film system by Ali et al. [12], along with the new hypotheses about EB being mainly driven by disordered interfacial spins as "*spin clusters*" at the interface between the FM and AFM layers as proposed by O'Grady et al. [6], or by magnetically hard particles (e.g., CoFe$_2$O$_4$) formed at the interface between the FeNi and CoO layers in FeNi/CoO thin films as suggested by Berkowitz et al. [11], are important clues for assessing the underlying mechanisms of EB in FM/AFM thin film systems; however, further experimental and theoretical studies are needed to verify these hypotheses. In a nanosystem composed of multiple magnetic phases/components or having multiple interfaces, it becomes more challenging to understand the magnetism of the system, due to complexity in magnetic interactions between phases of different degrees of magnetic ordering. For instance, our AC susceptibility measurements [114] have revealed a greater degree of magnetic frustration in the flower-like Au/Fe$_3$O$_4$ MNPs that have *multiple* interfaces, as compared to the dumbbell-like Au/Fe$_3$O$_4$ MNPs that possess only *single* interface between Au and Fe$_3$O$_4$. So, two important questions emerge and need to be addressed: *How does this magnetic frustration affect the low temperature magnetic behavior and EB in the flower-like Au/Fe$_3$O$_4$ MNPs?* and *Can it be decoupled from interfacial stress, surface and finite size effects?* In the case of FeO/Fe$_3$O$_4$ core/shell nanocubes, Hai et al. [115] have found the presence of Fe$_3$O$_4$ in both the shell and the core, forming a complicated magnetic structure. Most recently, Kaur et al. [116] have reported on the complexity in magnetism of watermelon-like iron nanoparticles, in which the large EB effects are ascribed to the AFM layer (Cr$_2$O$_3$) formed with the iron oxide shell, and the exchange interaction competes with the dipolar interaction with the increase of $\alpha$-FeCr grains in the Fe core.

From the application perspective, spintronics and biomedical applications of iron oxide-based nanoparticle systems are very promising. However, further efforts are needed to improve their structural stability and properties. A synergistic exploitation of the magnetic and photo-thermal properties of hybrid nanostructures such as Ag-Fe$_3$O$_4$ nanoflowers has recently revealed a new, effective approach for reducing the AC magnetic field and laser intensities in hyperthermia treatment [117]. This constitutes a key step towards optimizing the hyperthermia therapy through a combined multifunctional magnetic and photo-thermal treatment and improving our understanding of therapeutic process to specific applications that will entail coordinated efforts in physics, engineering, biology and medicine.

**Acknowledgments:** Research at the University of South Florida was supported by the U.S. Department of Energy, Office of Basic Energy Sciences, Division of Materials Sciences and Engineering under Award No. DE-FG02-07ER46438 (Synthesis and magnetic studies). Research at the University of Barcelona (Magnetic simulation) was supported by Spanish MINECO (MAT2012-33037, MAT2015-68772-P), Catalan DURSI (2009SGR856 and 2014SGR220), and European Union FEDER funds (Una manera de hacer Europa). Javier Alonso acknowledges the financial support provided through a postdoctoral fellowship from Basque Government.

**Conflicts of Interest:** The authors declare no conflict of interest.

## Appendix A

Atomistic Monte Carlo simulations are especially useful to elucidate the origin of the magnetic characteristics and the EB effect in MNPs systems. Essentially, with the Monte Carlo simulations we try to "follow" the evolution of the spins orientation as a function of, for example, the temperature of the system or the intensity of the applied magnetic field. During repetitive iterations, the orientation of the spins conforming the system is allowed to change, and every time a change would occur, the increase of the energy of the system (given by an appropriate Hamiltonian), $\Delta E$, is evaluated: if $\Delta E < 0$, the change in orientation is always allowed since it brings the system to a state of lower energy; but if $\Delta E > 0$, there is only a certain probability, given by the Boltzmann statistics $\exp(-\Delta E/k_B T)$, that the change in orientation will be allowed [118]. After a certain number of iterations, once the system has reached the equilibrium, the magnetization of the whole system will be recorded, the temperature or the field will be changed, and the process will be repeated again. This way we can simulate hysteresis loops or magnetization curves to demonstrate the peculiar magnetic behavior of the MNPs.

In these simulations, the magnetic ions are normally represented by either classical Ising spins, that can be in one of two states (+1 or −1), or Heisenberg spins, 3-dimensional vector *spins*, placed on the nodes of a real maghemite or magnetite structure sublattice in tetrahedral and octahedral coordination. The simulated MNPs can have very different radii and geometries, giving rise to different approaches in order to perform the Monte Carlo simulations. For example, in the case of a solid MNP, normally one has to distinguish between the spins on the surface and the spins in the inside of the nanoparticle, but for a simulated hollow morphology, one has both an outer and inner surface, with differentiated inner and outer surface spins, and the shell is also divided into different regions (crystallites), each one of them with an intrinsic random uniaxial anisotropy direction.

Despite this, in a first approximation, for all the Monte Carlo simulations mentioned in these article the same initial Hamiltonian can be employed:

$$\frac{H}{k_B} = - \sum_{\langle i,j \rangle} J_{ij}\left(\vec{S_i} \cdot \vec{S_j}\right) - \sum_i \vec{h} \cdot \vec{S_i} + E_{anis} \tag{A1}$$

The first term corresponds to the exchange interaction between nearest neighbours (nn): the exchange interaction depends on the coordination of the Fe ions (tetrahedric T or octahedric O) and the values correspond to real values. For example, for maghemite only $Fe^{3+}$ ions are present, and therefore we can consider $S = 1$ and $J_{ij}^{TT} = -21$ K, $J_{ij}^{OO} = -8.6$ K, and $J_{ij}^{TO} = -28.1$ K [61]. For magnetite, instead, we have to distinguish between $Fe^{2+}$ and $Fe^{3+}$ ions, thus if we assign a value of $S = 5/2$ for $Fe^{3+}$ and $S = 2$ for $Fe^{2+}$, then $J_{ij}^{TT} = -2.55$ K, $J_{ij}^{OO} = +14.6$ K, and $J_{ij}^{TO} = -67.7$ K [103]. The second term is the Zeeman energy, with $h = \mu H/k_B$ ($H$ is the magnetic field and $\mu$ corresponds to the magnetic moment of the ion, $Fe^{2+}$ or $Fe^{3+}$). And the third term is related to the magnetocrystalline anisotropy energy. For the anisotropy energy, as was commented before, normally one distinguishes between the surface spins, with reduced coordination and anisotropy constant $k_S$, and the core spins with full coordination and anisotropy constant $k_C$.

Therefore, the anisotropy energy term can be expressed as:

$$E_{anis} = k_S \sum_{i \in S} \sum_{j \in nn} \left(\vec{S_i} \cdot \hat{r_{ij}}\right)^2 - k_C \sum_{i \in C} \left(\vec{S_i} \cdot \vec{n_i}\right)^2 \tag{A2}$$

where $\hat{r_{ij}}$ is a unit vector joining spin $i$ with its nearest neighbors and $\hat{n_i}$ is a unit vector that depends on the specific morphology considered. For example, for a solid nanoparticle, is just a uniaxial vector, while for the hollow nanoparticle, $\widehat{n_t}$ is the random anisotropy axis for each one of the crystallites in which the simulation volume has been divided, as commented before. The values of $k_S$ and $k_C$ must be defined depending on the specific case, as indicated in the main text.

In certain cases, slight modifications to this Hamiltonian need to be introduced in order to take into account the specific characteristics of the system we are simulating. Other parameters such as the number of iterations, the field or temperature increments, etc. can be tuned in order to improve the accuracy of the simulated results while keeping the computation time required within a reasonable limit, making Monte Carlo simulations a robust and relatively easy to use tool to get a better depiction about the particular characteristics of the EB effect observed in the different systems analyzed in this article.

**Appendix B**

*Appendix B.1. Synthesis of Core/Shell Fe/γ-Fe₂O₃ and Hollow γ-Fe₂O₃*

The core/shell Fe/γ-Fe₂O₃ nanoparticles described in Section 4 were synthesized by thermal decomposition of organometallic compounds. A three necked flask was charged with Oleylamine, 70%, and 1-Octadecene, 90%, and the mixture was stirred at 140 °C under a mixture of 95% Ar + 5% H₂ gases for 2 h. The temperature was raised subsequently to 220 °C and Iron pentacarbonyl, Fe(CO₂)₅, was injected at 220 °C and refluxed for 20 min. A black precipitate followed by white smoke indicated the formation of nanoparticles. The sample cooled down to room temperature and a sufficient amount was removed for characterization. The result is core/shell Fe/γ-Fe₂O₃ nanoparticles. The formation of hollow nanoparticles is a consequence of Kirkendall effect which is a slow oxidation process. The rate of diffusion of iron is much faster than the rate of diffusion of oxygen, which will generate vacancies at the interface between the iron and the iron oxide. The super saturation of these vacancies will generate a void. At this point the sample alters from core/shell to core/void/shell. The core/void/shell sample will eventually morph to form a hole at the center of particle. After the initial attachment of the oxygen layer to the metal surface, the electrons from the core tunnel through the thin oxide layer at surface and ionize the oxygen leading to an electric field between the metal core and surface oxide layer. This electric field will subsequently drive the outward diffusion of ionized iron. A schematic of this process is depicted in Figure B1. To speed up the formation of hollow nanoparticles, the core/shell sample was annealed at 180 °C for one hour under a flow of oxygen. Core/shell and hollow nanoparticles were washed with a mixture of 3 mL Hexane, 95%, and 97 mL Ethanol, ≥99.5%.

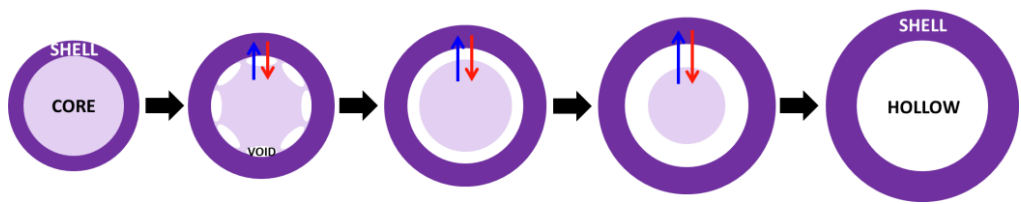

**Figure B1.** Schematic representation of the nanoparticles' conversion from core/shell, to core/void/shell and finally to hollow via the Kirkendall effect.

The bright field TEM images (see Figure B2) clearly show a lack of contrast, marking the presence of a cavity in the particle center, which is more obvious from the high resolution imaging. Further high resolution TEM imaging (HRTEM) shows that the particles are composed of randomly oriented grains of iron-oxide. The selected area electron diffraction matches well to FCC phase iron oxide.

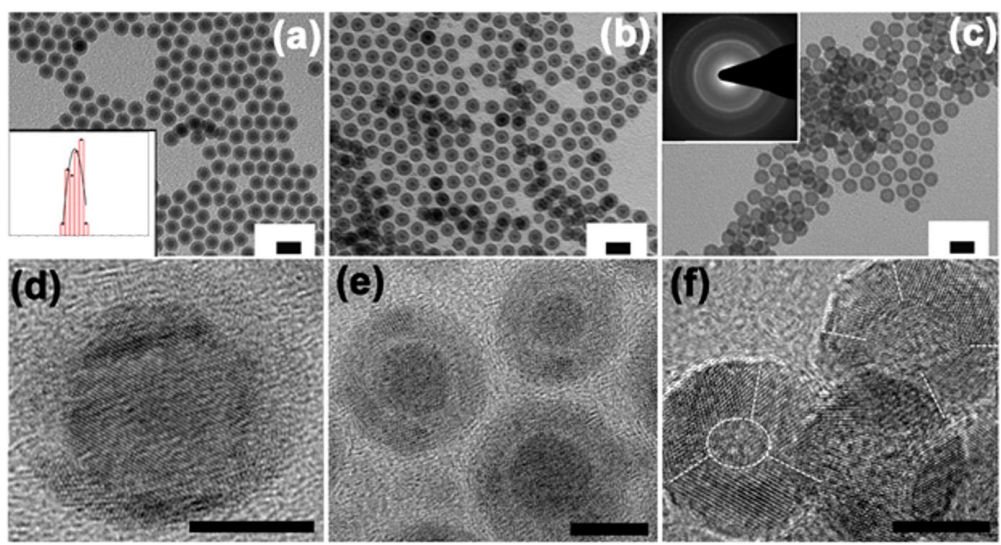

**Figure B2.** Bright-field TEM images of the 12 nm Fe/γ-Fe₂O₃ (**a**) core/shell, (**b**) core/void/shell, and (**c**) hollow nanoparticles; inset of (**a**) shows a histogram of the particle size populations for the 12 nm core/shell nanoparticles and inset of (**c**) shows SAED pattern of hollow nanoparticles. HRTEM images of (**d**) core/shell, (**e**) core/void/shell and (**f**) hollow nanoparticles. The scale bar is 20 nm in (**a–c**) and is 5 nm in (**d–f**). The discontinuous lines in (**f**) show grain boundaries of nanograins in the hollow nanoparticles.

*Appendix B.2. Synthesis of Ag/Fe₃O₄ Composite Nanoflowers*

Ag/Fe₃O₄ composite nanoflowers were synthesized using a one-step solvothermal process. In a typical reaction, 1.16 g of iron nitrate nonahydrate (Fe(NO₃)₃·9H₂O) was dissolved in 35 mL of ethylene glycol, followed by addition of 2.9 g of sodium acetate (NaAc) and 0.1 g of silver nitrate (AgNO₃). The solution was starred for 30 min to dissolve all of the reactants. The solution was then transferred to a 45 mL Teflon lined autoclave and heated at 200 °C for 24 h. After cooling the autoclave to room temperature, a black precipitate had formed. This black precipitate was washed several times with water and ethanol. The final product was dispersed in ethanol for further characterization. Different sizes of Ag/Fe₃O₄ composite nanoflowers were synthesized by varying the concentration of the precursors during the synthesis.